\documentclass[elsart12,epsf,eqsecnum]{revtex4}

\usepackage{epsfig,epsf}
\begin{document}











\def\beq{\begin{equation}}

\def\eeq{\end{equation}}
\def\eq#1{{Eq.~(\ref{#1})}}

\def\fig#1{{Fig.~\ref{#1}}}
             
\newcommand{\bas}{\bar{\alpha}_S}

\newcommand{\as}{\alpha_S} 

\newcommand{\bra}[1]{\langle #1 |}

\newcommand{\ket}[1]{|#1\rangle}

\newcommand{\bracket}[2]{\langle #1|#2\rangle}

\newcommand{\intp}[1]{\int \frac{d^4 #1}{(2\pi)^4}}

\newcommand{\mn}{{\mu\nu}}

\newcommand{\tr}{{\rm tr}}

\newcommand{\Tr}{{\rm Tr}}

\newcommand{\T} {\mbox{T}}
\newcommand{\W} {\text{W}}

\newcommand{\braket}[2]{\langle #1|#2\rangle}

\newcommand{\ab}{\bar{\alpha}_S}

\newcommand{\bea}{\begin{eqnarray}}

\newcommand{\eea}{\end{eqnarray}}

\setcounter{secnumdepth}{7}

\setcounter{tocdepth}{7}

\parskip=\itemsep               






\setlength{\textwidth}{168mm}





\newcommand{\beqar}[1]{\begin{eqnarray}\label{#1}}

\newcommand{\eeqar}{\end{eqnarray}}

\newcommand{\m}{\marginpar{*}}

\newcommand{\lash}[1]{\not\! #1 \,}

\newcommand{\nn}{\nonumber}

\newcommand{\D}{\partial}

\newcommand{\h}{\frac{1}{2}}

\newcommand{\g}{{\rm g}}

\newcommand{\el}{{\cal L}}

\newcommand{\A}{{\cal A}}

\newcommand{\Ka}{{\cal K}}

\newcommand{\al}{\alpha}

\newcommand{\be}{\beta}

\newcommand{\ep}{\varepsilon}

\newcommand{\ga}{\gamma}

\newcommand{\de}{\delta}

\newcommand{\De}{\Delta}

\newcommand{\et}{\eta}

\newcommand{\ka}{\vec{\kappa}}

\newcommand{\la}{\lambda}

\newcommand{\ph}{\varphi}

\newcommand{\si}{\sigma}

\newcommand{\ro}{\varrho}

\newcommand{\Ga}{\Gamma} 

\newcommand{\om}{\omega}

\newcommand{\La}{\Lambda}  

\newcommand{\tG}{\tilde{G}}

\renewcommand{\theequation}{\thesection.\arabic{equation}}



%

\def\ap#1#2#3{     {\it Ann. Phys. (NY) }{\bf #1} (19#2) #3}

\def\arnps#1#2#3{  {\it Ann. Rev. Nucl. Part. Sci. }{\bf #1} (19#2) #3}

\def\npb#1#2#3{    {\it Nucl. Phys. }{\bf B#1} (19#2) #3}

\def\plb#1#2#3{    {\it Phys. Lett. }{\bf B#1} (19#2) #3}

\def\prd#1#2#3{    {\it Phys. Rev. }{\bf D#1} (19#2) #3}

\def\prep#1#2#3{   {\it Phys. Rep. }{\bf #1} (19#2) #3}

\def\prl#1#2#3{    {\it Phys. Rev. Lett. }{\bf #1} (19#2) #3}

\def\ptp#1#2#3{    {\it Prog. Theor. Phys. }{\bf #1} (19#2) #3}

\def\rmp#1#2#3{    {\it Rev. Mod. Phys. }{\bf #1} (19#2) #3}

\def\zpc#1#2#3{    {\it Z. Phys. }{\bf C#1} (19#2) #3}

\def\mpla#1#2#3{   {\it Mod. Phys. Lett. }{\bf A#1} (19#2) #3}

\def\nc#1#2#3{     {\it Nuovo Cim. }{\bf #1} (19#2) #3}

\def\yf#1#2#3{     {\it Yad. Fiz. }{\bf #1} (19#2) #3}

\def\sjnp#1#2#3{   {\it Sov. J. Nucl. Phys. }{\bf #1} (19#2) #3}

\def\jetp#1#2#3{   {\it Sov. Phys. }{JETP }{\bf #1} (19#2) #3}

\def\jetpl#1#2#3{  {\it JETP Lett. }{\bf #1} (19#2) #3}


\def\ppsjnp#1#2#3{ {\it (Sov. J. Nucl. Phys. }{\bf #1} (19#2) #3}

\def\ppjetp#1#2#3{ {\it (Sov. Phys. JETP }{\bf #1} (19#2) #3}

\def\ppjetpl#1#2#3{{\it (JETP Lett. }{\bf #1} (19#2) #3} 

\def\zetf#1#2#3{   {\it Zh. ETF }{\bf #1}(19#2) #3}

\def\cmp#1#2#3{    {\it Comm. Math. Phys. }{\bf #1} (19#2) #3}

\def\cpc#1#2#3{    {\it Comp. Phys. Commun. }{\bf #1} (19#2) #3}

\def\dis#1#2{      {\it Dissertation, }{\sf #1 } 19#2}

\def\dip#1#2#3{    {\it Diplomarbeit, }{\sf #1 #2} 19#3 }

\def\ib#1#2#3{     {\it ibid. }{\bf #1} (19#2) #3}

\def\jpg#1#2#3{        {\it J. Phys}. {\bf G#1}#2#3}  

%


%


\def\thefootnote{\fnsymbol{footnote}} 

\title{
\Large  \bf Treading on the cut: Semi inclusive observables at high energy.}
\author{A. Kovner $^a$, M. Lublinsky $^a$ and H. Weigert $^b$} 

\address{$^a$Physics Department, University of Connecticut, \\ 2152 Hillside
Road, Storrs, CT 06269-3046, USA
\\
$^b$Department of Physics, 
The Ohio State University \\
Columbus, OH 43210, USA}

\begin{abstract}We develop the formalizm for calculating semi inclusive observables at high energy in the JIMWLK/KLWMIJ approach. This approach is valid for scattering of a small perturbative projectile on a large dense target. We consider several examples including diffractive processes, elastic and inclusive over the target degrees of freedom, scattering with fixed total transverse momentum transfer and inclusive gluon production. We discuss evolution of these observables with respect to various rapidity variables involved in their definitions (total rapidity, rapidity gap, width of diffractive interval etc.). We also discuss the dipole model limit  of these observables and derive closed simple (as opposed to functional) differential equations in this approximation. We point out that there are some interesting differences between the full JIMWLK/KLWMIJ evolution and the dipole model evolution of diffractive cross section.
In particular we show that in the dipole approximation the target does not diffract beyond the valence rapidity interval, consistently with the intuition suggested by the Pomeron fan diagramms. On the other hand such diffractive processes do exist in the full JIMWLK/KLWMIJ approach, albeit suppressed by the factor $1/N_c^2$.
\end{abstract}
\maketitle






\def\thefootnote{\arabic{footnote}} 

\section{Introduction and conclusion}
Attempts at understanding of high energy hadronic scattering have been at the
forefront of the QCD research for a very long time
\cite{BFKL,ELLA,BKP,GLR,MUQI,Mueller,3P,3P1,NP,BV,mv}. In recent years these have
increasingly focused on the physics of saturation
\cite{GLR},\cite{MUQI},\cite{mv} in the formal framework of what is being referred to
intermittently as the Balitsky-Kovchegov (BK) \cite{Balitsky},\cite{Kovchegov} or the JIMWLK
evolution equations\cite{JIMWLK},\cite{cgc}.  This set of equations describes
evolution of the scattering amplitude of an arbitrary perturbative hadronic
projectile on a dense hadronic target. The evolution of the projectile wave
function in this approach is perturbative, but the interaction with the target
is resummed to all orders in the target gluon field strength in the eikonal
approximation.  Recently a lot of thought has been devoted to finding the
extension of this approach to include nonlinear effects also in the projectile
wave function - the so called Pomeron loops \cite{Balitsky05,BraunPL,IM, shoshi1,  IMM, IT, MSW, LL3,  kl,  kl1,
 something, levin, blaizot,kl4, kl5, SMITH,Hatta,Kovchegov05,genyakozlov1,SX,yinyang}. It is also worth mentioning
that the JIMWLK/KLWMIJ equation and its generalization including the Pomeron
loops has been interpreted as the perturbative QCD incarnation of the Reggeon
field theory\cite{rft}, and some properties of its Hilbert space and spectrum
have been studied\cite{yinyang}.

The BK/JIMWLK evolution equation directly describes the evolution of the total
cross section. Indeed most of the work in this framework has concentrated on
understanding of this quantity. On the other hand, as is now clear, the total
cross section at asymptotically high energy cannot be reliably calculated in
this approach, since it violates the Froissart bound\cite{froissart}. On the
other hand one does expect that other more exclusive observables have a better
chance to be calculable in the perturbative saturation approach. It is thus
important to understand how to adapt the formalism to calculation of such semi
inclusive quantities. This is the main purpose of the present paper. We hasten
to note that several semi inclusive observables have been already discussed in
this or similar frameworks. In particular in the dipole model approach
diffractive dissociation cross section was considered in
\cite{Kovchegov:1999ji}, single gluon inclusive spectrum was calculated in
\cite{Kovchegov:2001sc} and double gluon emission spectrum was discussed in
\cite{JMK}.  Within the JIMWLK approach proper, the single gluon spectrum and
the double gluon spectrum with rapidities of the two gluons close to each
other has been calculated in \cite{Baier:2005dv} and the formalism for
calculating diffractive cross sections has been put forward in \cite{HWS}.

The aim of the present paper is to consistently and fully develop the formalism
for calculating semi-inclusive observables within the JIMWLK/KLWMIJ framework.
Many elements of the developments in this paper are independent of the
explicit form of the JIMWLK Hamiltonian and those generalize straightforwardly
to the as yet unknown complete high energy evolution Hamiltonian including
Pomeron loops.  However by using the JIMWLK Hamiltonian for the evolution of the target wave function 
we restrict the applicability of the current calculation to large target and small projectile. More specifically, the target already at zero rapidity is assumed to be a dense system, such as nucleus. This allows one to use JIMWLK evolution for the target wave function already at zero rapidity. The projectile on the other hand is assumed to be a dilute system not only at the initial rapidity but all the way to the total rapidity of the process $Y$. This allows one to use KLWMIJ evolution for the projectile wave function all the way up to the final rapidity $Y$. These conditions ensure that the "pomeron splitting" contributions to the evolution of the target wave function and  the "pomeron merging" contributions to the evolution of the projectile wave function are unimportant throughout the whole rapidity range considered. The calculation is not appropriate to "proton-proton" scattering at asymptotically high energies, but rather is pertinent to "DIS on a nucleus" at preasymptotic energies. We stress that this limitation is not specific for application of JIMWLK evolution to semiinclusive observables, but is inherent in the JIMWLK approach as a whole.

Our starting point is the energy evolution of the light cone
wave function \cite{kl1,zakopane}. We use essentially the same approach as in
\cite{Baier:2005dv} and develop it further to include a wider class of observables. We
consider several examples, in particular variety of elastic and/or diffractive
cross sections, cross section with fixed transverse momentum transfer and
inclusive gluon spectrum. In all these cases we define the appropriate
observable, derive its evolution with rapidity (total rapidity of the process
and/or width of the rapidity gap and/or width of the diffractive interval) and
discuss in detail the dipole model limit for each one of the observables. We
make an explicit link with the work \cite{HWS} and provide explicit
definitions of the basic quantities used in \cite{HWS} in terms of physical
$S$ matrix elements.

Apart from setting the unified framework for discussion of semi inclusive
observables, we make several specific points which are worth noting. First, in
addition to the diffractive cross section $N^D_{El}$ discussed in
\cite{Kovchegov:1999ji} which is elastic in the target degrees freedom, we
consider $N^D_D$ which is inclusive over the final states of the target within
a small target side rapidity interval. We confirm that the two quantities are
different, and derive the dipole limit evolution for both (equation for
$N^D_{El}$ being the Kovchegov-Levin equation \cite{Kovchegov:1999ji}). We
also show that the more general double diffractive cross section in the dipole approximation does not
depend on the width of the diffractive interval on the target side, but only
on the total distance in rapidity between the target and the diffractive
remnants of the projectile.  This property is straightforward in terms of the
pomeron fan diagrams, since once the final state cut enters the gap, it
cannot cross any more Pomerons, and thus the width of the diffractive interval
on the target side is always confined to a finite fixed value. 
 Interestingly, 
we show that in the full JIMWLK/KLWMIJ framework this property does not hold, 
and diffraction of the target is possible. Thus the $1/N_c$ corrections to the dipole model 
which are present in the JIMWLK equation lead to a qualitatively different behavior 
of the diffractive cross section. 

 Any experimental observation of double diffractive processes can be  viewed 
as a measure of physics beyond the fan diagram approximation. This physics is traditionally
associated with effects of pomeron loops, which have double diffractive processes
as experimental signature. We observe, however,  that  the JIMWLK/KLWMIJ framework beyond the dipole model limit
leads to similar signatures even though it does not take into account high density effects in the projectile wave function.

This paper is structured as follows. In Section 2 we recall the general
framework of the light cone wave function evolution and how it leads to the
JIMWLK/KLWMIJ evolution equation. We also discuss the large $N_C$ limit and
how it leads to the dipole model and its generalizations. In Section 3 we
introduce the method of calculating semi inclusive observables in this
framework and consider variety of elastic and diffractive observables and
derive their evolution equations with respect to total rapidity as well as
rapidity gap and diffractive rapidity intervals. In Section 4
we derive the dipole limit of the evolution equations in all the cases. This
is necessary for any possible numerical implementation of the evolution with
the view of phenomenological applications since it reduces the evolution
equations to simple rather than functional differential equations. In Sec. 5
and 6 we discuss the scattering cross section with total transverse momentum
transfer and the inclusive single gluon production, as examples of
non-diffractive observables which are calculated by the same methods.

\section{High energy scattering: the general setup}
 
\subsection{Evolution of the wavefunction}
In this section we review the approach to high energy evolution based directly on the evolution of a hadronic wave function. We will follow the formalism of \cite{urs}. We concentrate on the gluonic part of the wave function, although including quarks does not pose any additional differences.

Consider an arbitrary high energy hadron with large rapidity $Y_0$. Its wave function in the gluon Fock space can be written as
\begin{equation}
|v\rangle_{Y_0}\,=\,\Psi[a^{\dagger a}_{v,i}(x)]|0\rangle_{Y_0}
\label{wf}
\end{equation}
The gluon creation operators $a^\dagger$ depend on the transverse coordinate, and also on the longitudinal momentum $k^+$. 
 The gluon operators $a_v$ in eq.(\ref{wf}) all have longitudinal
 momenta above some cutoff $\Lambda$. We  refer to these degrees of freedom as "valence".
Henceforth we omit the dependence on longitudinal momentum in our expressions, as the momentum enters only as a spectator
 variable and only determines the total phase space available for the evolution.

When boosted, the valence wave function gets dressed by the evolution "Cloud" operator $\Omega$ \cite{inprep}. Physically this operator creates the soft Weiszacker-Williams gluon field which accompanies  the boosted valence gluons. It therefore depends
on the color charge density of the valence degrees of freedom ($\rho$) and also involves creation operators of softer modes  $a_{k^+}$ with longitudinal momenta below $\Lambda$.  The evolved wave function has the following structure 
\beq
|\Psi \rangle_Y\,=\, \Omega_Y(\rho,\,a)\,|v\rangle_{Y_0}\,. \ \ \ \ \ \ \ \ \ \ \  \ \ \ \ \ \ 
\eeq
The evolution operator $\Omega$ is explicitly known in the dilute limit ($\rho\rightarrow 0$) only. In this limit, for $Y-Y_0=\delta Y\ll 1$, it is 
 the coherent operator 
\beq\label{co}
\Omega_{\rho\rightarrow 0}=C_{\delta Y}\,=\,{\rm Exp}\left\{
i\int d^2xb_i^a(x)\int_{\Lambda}^{e^{\delta Y}\,\Lambda}{dk^+\over \pi^{1/2}| k^+|^{1/2}} 
\left[a^{ a}_i(k^+, x)\,+\,a^{\dagger a}_i(k^+, x))\right]\right\}\,.
\eeq
Here the creation operators $a^\dagger(k^+)$ create gluons with soft momenta, which are not present in the valence
 state $|v\rangle$. The field $b$ depends only on the valence degrees of freedom through
 \begin{equation}
b^a_i(z)\,=\,{g\over 2\pi}\int d^2x{(z-x)_i\over (z-x)^2}\,\rho^a(x)\,.
\end{equation} 
For a finite evolution interval the appropriate expression is
\beq\label{co1}
\Omega_Y=C_Y\,=\,{\cal P}{\rm Exp}\left\{
i\int_{\Lambda}^{e^{Y-Y_0}\,\Lambda}\int d^2xb_i^a(x,k^+){dk^+\over \pi^{1/2}| k^+|^{1/2}} 
\left[a^{ a}_i(k^+, x)\,+\,a^{\dagger a}_i(k^+, x))\right]\right\}\,.
\eeq
where ${\cal P}$ denotes the path ordering with respect to $k^+$, and the $k^+$  dependent field $b$ includes the charge density of all modes harder than $k^+$
\begin{equation}
b^a_i(z)\,=\,{g\over 2\pi}\int d^2x{(z-x)_i\over (z-x)^2}\,[\rho^a(x)+\int_{\Lambda}^{k^+}dp^+a^\dagger(p^+)T^aa(p^+)]\,.
\end{equation} 
One can resum some corrections away from the low charge density limit by keeping the full nonlinear relation between the Weizsacker-Williams field and the color charge density\cite{kl1, SMITH}. In this case $b^a_i$ is determined as the solution of
 the "classical" equation of motion 
\begin{eqnarray}
&&\partial_ib_i^a\,+\,g\,\epsilon^{abc}\,b^b_i(x)\,b^c_i(x)\,=\,g\,\rho^a(x)\nonumber\\
&&\epsilon_{ij}\,[\partial_ib^a_j\,-\,\partial_jb^a_i\,+\,g\epsilon^{abc}b^b_i\,b^c_j]\,=\,0
\label{b}
\end{eqnarray}
The coherent operator $C$ dresses the valence wave function by the cloud of the Weizsacker-Williams gluons:
\begin{equation}\label{shift}
C^{\dagger}\, A^a_i(k^+, x)\,C\,=\,A^a_i(k^+, x)+{i\over k^+}b_i^a(x)\,.
\end{equation}

Given this evolution of the hadronic wave function one can calculate the evolution of an arbitrary  observable $\hat{\cal O}(\rho)$ which depends on the color charge density. For example, the $S$ matrix in eikonal approximation belongs to this class of observables. In the by now standard notation 
\begin{equation}
\langle v|\,\hat{\cal O}[\rho]\,|v\rangle\, =\,\int D\rho \,W[\rho]\,\,{\cal O}[\rho]\,.
\end{equation}
As discussed in detail in \cite{kl}, the integration variable $\rho$ on the
right hand side of this equation is understood to depend on transverse
coordinates as well as an additional coordinate $x^-$, which can be thought of 
either as the longitudinal coordinate of the hadronic wave function or as a mathematical
``ordering coordinate'' present to enforce correct commutation relations of the
operators $\hat\rho^a$.

The evolution of the expectation value is then given by
\beq
\frac{d\,\langle v|\hat{\cal O}|v\rangle}{d\,Y}\,=\,\lim_{{Y\rightarrow Y_0}}
\frac{\langle v|\Omega^\dagger_Y\, \hat{\cal O}(\rho+\delta\rho)\,\Omega_Y|v\rangle\,\,
-\,\,\langle v|\hat{\cal O}(\rho)|v\rangle}{Y\,-\,Y_0}\,=\,-\,\int D\rho\, W[\rho]\,H[\rho]\,\,{\cal O}[\rho]\,.
\label{diag}
\eeq

The color charge density in the first term contains also the contribution of the soft gluons generated by the evolution
\begin{equation}
\delta\rho^a(x)\,=\,\int^{e^{Y-Y_0}\,\Lambda}_{\,\Lambda} dk^+\,a^{\dagger b}_i(k^+,x)\,T^a_{bc}\,a^{c}_i(k^+,x)
\end{equation}
where $T^a_{bc}=if^{abc}$ is the $SU(N)$ generator in the adjoint representation. 
This is conveniently represented in terms of the charge density shift operator (which also has the meaning of the ``dual'' to the Wilson line operator)
\begin{equation}
R(z)^{ab}\,\,=\,\,
\left[{\cal P}\exp{\int_{0}^{1} d z^-\,T^c\,{\delta\over\delta\rho^c(z,\,z^-)}}\right]^{ab}
\label{rr}
\end{equation}
The evolution Hamiltonian $H$ generally can be written as
\beq\label{chi}
H[\rho,{\delta\over\delta\rho}]\,=-\,\frac{d}{dY}\,\langle 0_a|\Omega^\dagger_Y(\rho,a)\,
\Omega_Y(R\,\rho, R\,a)|0_a\rangle{|_{Y=Y_0}}\,=\,\,\sum_n\,Q^\dagger_n[R]\,Q_n[R]\,.
\eeq
The last equality in (\ref{chi}) is given in terms of the $n$-particle production amplitudes $Q_n$ \cite{yinyang}.
Each $Q_n$ depends on $n$ transverse coordinates, color and Lorentz indices which we do
 not indicate explicitly.
 This expression is formally valid for arbitrary charge density. To write down an explicit expression for $H$ we specify to the dilute regime. In this case only one soft gluon is created in one step of the evolution and  the Hamiltonian gets contribution only from the one gluon ($n=1$) production amplitude:
\beq
H^{KLWMIJ}[\rho,{\delta\over\delta\rho}]\,=-\,\frac{d}{dY}\,
\langle 0_a|C^{\dagger}_Y(\rho,a)\,C_Y(R\,\rho, R\,a)|0_a\rangle|_{Y=Y_0}\,=\,
\int{d^2z \over 2\,\pi}\, Q^a_i(z)\,Q^a_i(z)
\end{equation}
where the one gluon emission amplitude $Q^a_i(z)$ is defined as (we suppress the index $n=1$)
\begin{equation}
\label{Q}
 Q^a_i(z)\,=\,R^{ab}(z)\,b^b_{L\,i}(z)\,-\,b^a_{R\,i}(z)
\end{equation}
Here the gluon field $b$ is the function of the $SU(N_C)$ rotation generators
\begin{equation}
b^a_{i,L(R)}(z)\,=\,{g\over 2\pi}\,\int d^2x{(z-x)_i\over (z-x)^2}\,J_{L(R)}^a(x)\,.
\end{equation} 
with
\begin{equation}
J_R^a(x)=-{\rm tr} \left\{R(x)T^{a}{\delta\over \delta R^\dagger(x)}\right\}, \ \ \ \  J_L^a(x)=
-{\rm tr} \left\{T^{a}R(x){\delta\over \delta R^\dagger(x)}\right\}, \ \ \ \  \ \ \ \ \ \ \ 
J_L^a(x)\,\,=\,\,[R(x)\, J_R(x)]^a.
\end{equation}
so that 
\begin{eqnarray}\label{KL}
Q^a_i(z)\,=\,{g\over 2\pi}\,\int\, d^2x{(x-z)_i\over (x-z)^2}\,[R^{ab}(z)\,-\,R^{ab}(x)]\, J^b_L(x)\,.
\end{eqnarray}
We refer to the evolution Hamiltonian in this dilute limit as the KLWMIJ Hamiltonian \cite{kl}.

The partial resummation of the nonlinearities mentioned above which corresponds to
keeping $b_{L,R}$ as solutions of the full classical equations of motion with
sources given by the generators of the left/right color rotations $J_{L,R}$\cite{kl1,
  SMITH} that leads to the evolution Hamiltonian $H^{KLWMIJ+}$.

We note that for the derivation of eq.(\ref{diag}) it was not crucial to consider a diagonal matrix element of the operator ${\cal O}$. 
The same derivation can be repeated straightforwardly for a generic non-diagonal matrix element as well.
Defining 
\begin{equation}
\langle v|\,\hat{\cal O}[\rho]\,|v'\rangle \,=\,\int D\rho\, \tilde W[\rho]\,\,{\cal O}[\rho]
\end{equation}
 we find the evolution
\begin{eqnarray}
\frac{d\,\langle v|\hat{\cal O}|v'\rangle}{d\,Y}\,&=&\,\lim_{{Y\rightarrow Y_0}}
\frac{\langle v|\Omega^\dagger_Y\, \hat{\cal O}(\rho+\delta\rho)\,\Omega_Y|v'\rangle\,\,
-\,\,\langle v|\hat{\cal O}(\rho)|v'\rangle}{Y\,-\,Y_0}\,=-\,\langle v|\,H[\rho]\,\,\,\hat{\cal O}(\rho)\,|v'\rangle\nonumber\\
&=&-\,\int D\rho \,\tilde W[\rho]\,H[\rho]\,\,{\cal O}[\rho]
\end{eqnarray}
with the same Hamiltonian as in eq.(\ref{chi}).
We will need to use this fact in the following.

\subsection{High energy scattering}

Throughout this paper we treat the scattering of  fast gluons of the projectile on the target in eikonal approximation.
We denote a S-matrix of a single gluon which scatters on a fixed configuration of chromoelectric field of the target by 
\begin{equation}
S^{ab}(x)=\langle 0| a^a_i(x)\,\hat S\,a^{\dagger b}_i(x)|0\rangle
\label{sgluon}
\end{equation}
where $\hat S$ is the second quantized $S$-matrix operator of the field theory which in the eikonal approximation\footnote{This expression is equivalent to eqs.(2.4-2.5) of \cite{kl4}.}
 is
\beq
\hat S\,=\,\exp\left\{i\int d^2x\,\hat\rho_P^a(x)\,\hat\alpha_T^a(x)\right\}\,.
\eeq
In the natural projectile light cone gauge ($A^-=0$) the large target field component is $A^+$, which we denote by
$\alpha_T$. 
The single gluon S-matrix $S^{ab}$ does not depend on the polarization of the gluon and is diagonal in the transverse
coordinate $x$ and is given by 
\begin{eqnarray}
\label{S}
S(x)\,\,=\,\,{\cal P}\,\exp\{i\int_{0}^{1} dy^-\,T^a\,\alpha^a_T(x,y^-)\}\,.
\end{eqnarray}

For a composite projectile which has some distribution of gluons in its wave function
 the eikonal $S$-matrix can be written in the form analogous to 
$S(x)$, see \cite{something}
\begin{equation}
\Sigma^{PP}[\alpha_T]\,\equiv\,\langle P|\,\hat S\,|P\rangle\,=\,\int d\rho_P\,\,W^P[\rho_P]\,\,
\,\exp\left\{i\int_{0}^{1} dy^-\int d^2x\,\rho_P^a(x,y^-)\,\alpha_T^a(x,y^-)\right\}
\label{s1}
\end{equation}
with $x_i$ - transverse coordinate.  The operator $\hat\rho_P(x_i)$ is the
color charge density in the projectile wave function at a given transverse
position, while $W^P[\rho]$ is the same weight functional as appearing in
eq.(\ref{diag}).  For a single gluon $\rho^a(x_i)=T^a \delta^2(x_i-x^0_i)$,
and eq.(\ref{s1}) reduces to eq.(\ref{S}).

To obtain the total $S$-matrix of the scattering process at a given rapidity
$Y$ one has to average $\Sigma$ of eq.(\ref{s1}) over the distribution of the
color fields in the target. Denoting the probability density to find a
particular configuration $\alpha_T(x,x^-)$ by $W^T[\alpha_T(x,x^-)]$ we write
\begin{equation}
{\cal S}(Y)\,=\,\int\, D\alpha_T^{a}\,\, 
W^T_{Y_0}[\alpha_T(x,x^-)]\,\,\Sigma^{PP}_{Y-Y_0}[\alpha_T(x,x^-)]\,.
\label{ss}
\end{equation}
In \eq{ss} we have chosen the frame where the target has rapidity $Y_0$ while
the projectile carries the rest of the total rapidity $Y-Y_0$.  Lorentz
invariance requires ${\cal S}$ to be independent of $Y_0$.

The high energy evolution of the $S$-matrix follows from
eqs.(\ref{diag},\ref{s1}) as
\begin{equation}
\frac{d}{d\,Y}\,{\cal S}\,=-\,\int\, D\alpha_T^{a}\,\, W^T_{Y_0}[\alpha_T(x,x^-)]\,\,\,
H\left[\alpha_T,\frac{\delta}{\delta\,\alpha_T}\right]\,\,\,
\Sigma^{PP}_{Y-Y_0}[\alpha_T(x,x^-)]\,.
\label{hee}
\end{equation}
with the Hamitonian eq.(\ref{chi}) with $\rho$ substituted by
${i\delta\over\delta\alpha^a(x,x^-)}$ The Hamiltonian can be viewed as acting
either to the right on $\Sigma$ or to the left (as it is Hermitian) on $\W$:
\begin{equation}
{\partial\over\partial Y}\,\Sigma^{PP}\,\,=-\,\,H\left[\alpha_T,\,{\delta\over\delta\alpha_T}\right]\,
\,\Sigma^{PP}[\alpha_T]\,;
\ \ \ \ \ \ \ \ \ 
{\partial\over\partial Y}\,W^T\,\,=-\,\,
H\left[\alpha_T,\,{\delta\over\delta\alpha_T}\right]\,\,W^T[\alpha_T]\,.
\label{dsigma}
\end{equation} 
As was shown in \cite{something} in order for the total $S$-matrix to be
Lorentz invariant and symmetric between the projectile and the target, the
evolution Hamiltonian $H$ must be self dual\cite{Balitsky05}. That is it has
to be invariant under the Dense-Dilute Duality transformation
\begin{equation}\label{duality}
\alpha^a(x,x^-)\,\rightarrow \,i\,{\delta\over\delta\rho^a(x,x^-)},
\ \ \ \ \ \ \ \ \ \ \ \ \ \ \ \, 
{\delta\over\delta\alpha^a(x,x^-)}\,\rightarrow\, -i\,\rho^a(x,x^-),
\ \ \ \ \ \ \ \ \ \ \ \ \ \ \ \, S\,\rightarrow R
\end{equation}
Hence
\begin{equation}
H\left[\alpha,\,{\delta\over\delta\alpha}\right]\,=\,\sum_n\,Q^\dagger_n[S]\,Q_n[S]\,.
\end{equation}
In the situation where the target is large and the projectile is small the
symmetry between the target and the projectile is irrelevant. In this limit
the Hamiltonian is given by the JIMWLK expression \cite{JIMWLK,cgc} which is
the dual of the $H^{KLWMIJ}$
\begin{eqnarray}\label{JIMWLK}
H^{JIMWLK}\,=\,\int_z\,\,  Q^a_i(z)\, Q^a_i(z)
\end{eqnarray}
with $ Q[S]$ obtained from $Q[R]$ (\ref{Q}) by substitution $S$ for $R$.

Some of the derivations in this paper are independent of the explicit form of the Hamiltonian. However whenever the explicit form is required we are going to use the JIMWLK
Hamiltonian eq.(\ref{JIMWLK}) with
\begin{eqnarray}
\label{qus}
Q^a_i(z,[S])\,=\,{g\over 2\pi}\,\int\, d^2x{(x-z)_i\over (x-z)^2}\,[S^{ab}(z)\,-\,S^{ab}(x)]\, J^b_L(x)
\end{eqnarray}
where
\begin{equation}\label{gen}
J_R^a(x)=-{\rm tr} \left\{S(x)T^{a}{\delta\over \delta S^\dagger(x)}\right\}, \ \ \ \  J_L^a(x)=
-{\rm tr} \left\{T^{a}S(x){\delta\over \delta S^\dagger(x)}\right\}, \ \ \ \  \ \ \ \ \ \ \ 
J_L^a(x)\,\,=\,\,[S(x)\, J_R(x)]^a.
\end{equation}
We record here two properties of this Hamiltonian which will be useful in our discussion of the evolution of diffractive observables.
For any function $F$ 
\begin{eqnarray}\label{property}
\int_z\left[Q^a_i(z,[S])\,+\,Q^a_i(z,[\bar S])\right]^2\,\,F[S\bar S^\dagger]&=&\int_z\left[Q^a_i(z,[S\bar S])\right]^2\,\,F[S\bar S]\nonumber\\
\left\{\int_z\left[Q^a_i(z,[S])\,+\,Q^a_i(z,[\bar S])\right]^2\,\,F[S,\, \bar S]\right\}_{S=\bar S}&=&\int_z\left[Q^a_i(z,[S])\right]^2
\,\,\left\{F[S,\, \bar S]\right\}_{S=\bar S}\,.
\end{eqnarray}
where $S$ and $\bar S$ are both arbitrary unitary matrices.

\subsection{Large $N_c$ lore: dipoles, quadrupoles  and such.} 
The general setup for the high energy evolution is that of functional
evolution equations for the scattering amplitudes, or equivalently effective
quantum field theory for the unitary matrix $S$. Some aspects of this Reggeon
field theory have been studied in \cite{rft},\cite{yinyang}. The problem of
its solution is however a formidable one, even though the theory is
considerably simple than full QCD. It is thus desirable to have a simple
truncated version of the theory which would reduce the complexity below the
level of quantum field theory. Such a truncation is offered by the formal
large $N_c$ limit and its simplest variant is Mueller's dipole
model\cite{Mueller, Mueller+}.

Any physical projectile must be color singlet, and the simplest color singlet
state is a fundamental dipole.  Assume for a moment that the projectile wave
function contains only dipoles as in  Mueller's dipole
model\cite{Mueller}. How does the scattering matrix of such a projectile
evolve? In the large $N_c$ limit color singlet objects evolve independently of
each other. Thus every dipole in the wave function leads his independent life.
Formally this means the following. The scattering matrix of a single dipole is
\begin{equation} 
   s(x,y)= {1\over N_C}tr[S_F(x)\, S^\dagger_F(y)] 
\end{equation}
where the subscript
$F$ denotes fundamental representation. The $S$ matrix of a dipole projectile
is therefore some function of the variable $s$ only 
\begin{equation}
\Sigma^{PP}[S]=\Sigma^{PP}[s] \,.
\end{equation}
For a given projectile wave function the
$S$-matrix $\Sigma[s]$ is easily calculated as 
\begin{equation} \Sigma^{PP}[s]=\sum_n
P_n\{(x_1,y_1),...,(x_n,y_n)\}s(x_1,y_1)...s(x_n,y_n)
\end{equation}
where $P_n\{(x_1,y_1),...,(x_n,y_n)\}$ is the probability to find $n$ dipoles
at the specified points in the incoming projectile wave function.

The high energy evolution of such a wave function in the large $N_c$ limit obeys Mueller`s dipole evolution 
re-expressed as Hamiltonian evolution in \cite{LL1}. The same result is derived starting directly from the 
JIMWLK equation \cite{kl1}. The result is
\begin{equation}
\label{dipoleev}
{d\Sigma_Y^{PP}[s]\over dY}\,=-\,H^{\text{dipole}}\,\Sigma^{PP}[s]
\end{equation}
with
\begin{eqnarray}
 \label{chidip}
H^{\text{dipole}}\,&=&
\,\frac{ \bar{\alpha}_s}{2\,\pi}\,
\int_{x,y,z}M_{x,y,z}\,\left[s(x,y)-s(x, z)\,s(y,z)\right]
\frac{\delta}{\delta s(x, y)} ;
\end{eqnarray}
with the dipole kernel
\begin{equation}
M_{x,y,z}\,= \,\frac{(x-y)^2}{(x-z)^2\,(z-y)^2}\,.
\end{equation}
The solution of the dipole evolution equation eq.~\eq{dipoleev} can be
expressed in terms of the solution of a simple differential
equation~\cite{LL1}
\begin{equation}
\Sigma_Y^{PP}[s]\,=\,\Sigma_{Y_0}^{PP}[s_Y] 
\end{equation}
where $s_Y$ solves the  BK equation (see Refs. \cite{BKT,BKN,MT,MP,kozlov} 
for analytical and numerical studies of the BK equation)
\begin{equation}
 \label{kovchegov}
 \frac{d s_Y}{dY}\,=\,\frac{ \bar{\alpha}_s}{2\,\pi}\,\int_z M_{x,y,z} 
\,[ s_Y(x,z)\, s_Y(z,y)\,-\, s_Y(x,y)]
\end{equation}
with the initial condition
\begin{equation}
s_{Y_0}\,=\,s
\end{equation}
At the same time the expression for the probability density,~\eq{ss},
turns into
\begin{equation}
{\cal S}^{\text{dipole}}(Y)\,=\,
\int\, Ds\,\, W^T_{Y_0}[s]\,\,\Sigma^{PP}_{Y-Y_0}[s]\, ,
\label{ss-dipole}
\end{equation}
i.e. the average over the gluon field of the target is rendered as an average
over an ensemble of dipoles at the initial $Y_0$ used in~\eq{kovchegov}.
This average still allows to accommodate nontrivial, non-factorized multi-s
correlators $\langle s(x_1,y_1)\cdots
s(x_n,y_n)\rangle_T$, see~\cite{LL1,LL2,JP,kl1,nestor}.

Further simplification is achieved if one assumes that the dipoles scatter on
the target independently. This amounts to factorization of the target averages of
the  dipole $s$-matrices
\begin{equation}
\langle s(x,y)\, s(u,v)\rangle_T
\,=\, \langle s(x,y)\rangle_T\langle s(u,v)\rangle_T
\end{equation} 
With this assumption, one replaces the ensemble average over target fields, or
alternatively over the ensemble of functions $s_{Y_0}(x,y)$ shown
in~(\ref{ss-dipole}), with a fixed initial function $s_{Y_0}(x,y)$.  We will
refer to this factorization property as the target mean field approximation.
Within the target mean field approximation
\begin{equation}
\langle \Sigma^{PP}_0[ s_Y ]\rangle_T\,=\,\Sigma^{PP}_0[\langle s_Y\rangle_T ]
\ .
\end{equation}
We stress that this mean field approximation does not follow logically from
the dipole model approximation for the evolution kernel eq.(\ref{dipoleev}),
but rather is an additional assumption about the properties of the target.
Physically it means that the target fields are assumed to fluctuate so
strongly that they are completely uncorrelated in different points in space.
In practice this is good assumption if the points are separated by a distance
larger than the saturation length $R_s=Q_s^{-1}$, which is also the
correlation length of the target fields. However for two dipoles separated by
a distance smaller than $R_s$ in the impact parameter space this approximation
is bound to break down. We will come back to this point later. For a more
detailed discussion see~\cite{LL1}.

The dipole model provides the simplest known framework for model discussion of high energy evolution. However as we shall see below, for some observables it is not sufficient to consider dipole degrees of freedom alone. In this case we have to allow for existence of quadrupoles.
Fortunately the large $N_c$ approach generalizes effortlessly 
beyond the dipole model. It is not necessary to assume that the projectile contains only dipoles. One can indeed allow the function $\Sigma^{PP}$ to depend also on the quadrupole degree of freedom
\beq q(x,y,u,v)\,=\,{1\over N_C}tr[S_F(x)\, S^\dagger_F(y)S_F(u)\, S^\dagger_F(v)]\,.
\eeq
The crucial property of the large $N_C$ evolution is that all singlets evolve independently. It is then straightforward to show that the evolution of a function $\Sigma^{PP}[s,q]$ in the large $N_C$ limit is
\begin{equation}
\label{quadrupoleev}
{d\Sigma_Y^{PP}[s,q]\over dY}\,=
-\left(\,H^{\text{dipole}}+H^{\text{quadrupole}}\right)\,\Sigma^{PP}_Y[s,q]
\end{equation}
with $H^{dipole}$ given by eq.(\ref{dipoleev}) and 
\begin{eqnarray}
\label{quadr}
H^{\text{quadrupole}}&=&\\
&=& \frac{\bar \alpha_s}{2\,\pi}\,\int_{x,y,u,v,z} \left\{-\,[M_{x,y;z}\,+\,M_{u,v;z}\,-\,L_{x,u,v,y;z}]
\,\,q_{x,y,u,v}\,-\,
 L_{x,y,u,v;z}\,s_{x,v}\,s_{y,u}\,-\right.\,\nonumber \\
&-&\,\,L_{x,v,u,y;z}\,s_{x,y}\,s_{u,v}\,+\,L_{x,v,u,v;z}\,q_{x,y,u,z}\,s_{z,v}
+\,L_{x,y,x,v;z}\,q_{z,y,u,v}\,s_{x,z}
\,+\nonumber \\
&+&\left.\,L_{x,y,u,y;z}\,q_{x,z,u,v}\,s_{z,y}\,+ \,L_{u,y,u,v;z}\,q_{x,y,z,v}\,s_{u,z}\right\} \times \ {\delta\over \delta q(x,y,u,v)}
\end{eqnarray}
where
\beq
L_{x,y,u,v;z}\,=\,\left[{(x\,-\,z)_i\over (x\,-\,z)^2}\,-\,{(y\,-\,z)_i\over (y\,-\,z)^2}\right ]\,\,
\left[{(u\,-\,z)_i\over (u\,-\,z)^2}\,-\,{(v\,-\,z)_i\over (v\,-\,z)^2}\right ]
\eeq
Again, the solution for the evolution equation eq.(\ref{quadrupoleev}) reduces to solution of ordinary (as opposed to functional) equations
\beq
\Sigma^{PP}_Y[s,q]=\Sigma^{PP}_0[s_Y, q_Y]
\eeq
where the dipole $s_Y$ solves the Kovchegov equation eq.(\ref{kovchegov})
and the quadrupole solves an analogous quadrupole evolution equation (see Appendix A and Ref. \cite{JMK})
\begin{eqnarray}\label{sq}
{d q_{x,y,u,v}\over d Y}&=&\frac{\bar \alpha_s}{2\,\pi}\,\int_z \left\{-\,[M_{x,y;z}\,+\,M_{u,v;z}\,-\,L_{x,u,v,y;z}]
\,q_{x,y,u,v}\,-\,
 L_{x,y,u,v;z}\,s_{x,v}\,s_{y,u}\,-\,L_{x,v,u,y;z}\,s_{x,y}\,s_{u,v}+\right.\,\nonumber \\
&+&\,L_{x,v,u,v;z}\,q_{x,y,u,z}\,s_{z,v}
\left.+\,L_{x,y,x,v;z}\,q_{z,y,u,v}\,s_{x,z}
\,+\,L_{x,y,u,y;z}\,q_{x,z,u,v}\,s_{z,y}\,+\,L_{u,y,u,v;z}\,q_{x,y,z,v}\,s_{u,z}\right\}
\,\nonumber \\
&&
\end{eqnarray} 
with the initial conditions
\beq s_{Y=0}\,=\,s;\ \ \ \ \ \ \ \ \  \ \ \ \ \ \ \ \ q_{Y=0}\,=\,q\,.
\eeq

In principle for this mixed dipole-quadrupole model one can again apply the target mean field approximation by assuming factorization of all the singlet averages
\begin{eqnarray}
&&\langle s(x,y)\, s(u,v)\rangle_T\,=\, \langle s(x,y)\rangle_T\,\langle s(u,v)\rangle_T\,; \\
&&\langle q(x,y,u,v)\, s(z,\bar z)\rangle_T\,=\, \langle q(x,y,u,v)\rangle_T\,\langle s(z,\bar z)\rangle_T\,;\nonumber \\
&&\langle q(x,y,u,v)\, q(\bar x,\bar y, \bar u,\bar v)\rangle_T\,=\, \langle q(x,y,u,v)\rangle_T\,
\langle q(\bar x, \bar y, \bar u,\bar v)\rangle_T\,.\nonumber
\end{eqnarray}
Generalization to higher multipoles is in principle straightforward, but we will not need for the observables considered in this paper.

\section{Semi-inclusive reactions}
\subsection{Generalities}

We are interested in calculating characteristics of the final states emerging
after a collision of the projectile, which at the initial rapidity has the
wave function $|P_v\rangle$ and the target with the wave function
$|T_v\rangle$. The target and the projectile are boosted before the collision
to the total rapidity $Y$.

Let $\Omega^P_{Y-Y_0}$ and $\Omega^T_{Y_0}$ denote the evolution operators for
projectile and target boosted to the rapidity $Y-Y_0$ and $Y_0$ respectively.
The time of interaction is set to be $t=0$.

The total wave function coming into the collision region is therefore at time
$t=0$
\begin{equation}
|\Psi_{in}\rangle\,=\,\Omega^{P}_{Y-Y_0}\,\Omega^{T}_{Y_0}\,
|P_v\rangle\,\vert T_v \rangle\,.
\end{equation}
The system emerges from the collision region with the wave function
\begin{equation}
|\Psi_{out}\rangle\,=\,\hat S\,\Omega^{P}_{Y-Y_0}\,\Omega^{T}_{Y_0}\,
|P_v\rangle\,\vert T_v \rangle\,.
\end{equation}
The system keeps evolving after the collision to the asymptotic time
$t\rightarrow +\infty$, at which point the measurement of an observable
$\hat{\cal O}$ is made.  As explained in \cite{Baier:2005dv} the evolution of
the outgoing wavefunction from $t=0$ to $t\rightarrow +\infty$ is given by the
Hermitian conjugate of the same operator $\Omega$. Thus the general setup for
computing any observable $\hat{\cal O}$ in the final state is
\begin{equation}\label{rdo0}
\langle \hat{\cal O}\rangle\,=\,\langle T_v\vert\,\langle\,P_v|\,
 \Omega^{P\, \dagger}_{Y-Y_0}\,
 \Omega^{T\, \dagger}_{Y_0}\,(1\,-\,\hat S^\dagger)\,
\Omega^{P}_{Y-Y_0}\,\Omega^{T}_{Y_0}\,\,\,\hat{\cal O}\,\,\,
\Omega^{P\, \dagger}_{Y-Y_0}\,\Omega^{T\, \dagger}_{Y_0}
\,(1\,-\,\hat S)\,\Omega^{P}_{Y-Y_0}\,\Omega^{T}_{Y_0}\,
|P_v\rangle\,\vert T_v \rangle
\end{equation}
where the factor $S-1$ ensures the proper subtraction of the contribution of
the initial state.  Generically, the observable ${\cal O}$ may depend both on
target and projectile valence degrees of freedom as well as gluon degrees of
freedom at intermediate rapidities.  An example of rapidity dependent
observables is diffraction considered in the next subsection (for example eq.(\ref{difel}) and single inclusive
gluon production discussed in Sect 6.

To express this in terms of scattering amplitudes we insert the resolution of identity on the projectile and target Hilbert spaces
$1\,=\,|p'\rangle\, \langle p'|$ and $1\,=\,|t'\rangle\,\langle t'|$.
Then we have
\begin{equation}\label{rdo}
\langle \hat{\cal O}\rangle\,=\,
\langle T_{Y_0}|\,\delta^{Pp'}\,-\,\hat\Sigma_{Y-Y_0}^{\dagger\, Pp'}
\,|t'\rangle\,\,\,{\cal O}^{p'p''}_{t't''}\,\,\,
\langle t''|\,\delta^{p''P}\,-\,\hat\Sigma_{Y-Y_0}^{p''P}\,|T_{Y_0}\rangle
\end{equation}
where we introduce a non-diagonal S-matrix element which is an operator acting on the target`s Hilbert space
\begin{equation}
\hat\Sigma_{Y-Y_0}^{Pp'}\,=\,\langle P_{Y-Y_0}|
\,\,\hat S\,\,|p'\rangle\,; 
\end{equation}
and the matrix element for the observable
\begin{equation}
{\cal O}^{p'p''}_{t't''}\,=\,\langle t'\vert\,\langle\,p'|
\Omega^{P}_{Y-Y_0}\,\Omega^{T}_{Y_0}\,\,\,\hat{\cal O}\,\,\,
\Omega^{P\, \dagger}_{Y-Y_0}\,\Omega^{T\, \dagger}_{Y_0}\,
|p''\rangle\,\vert t''\rangle\,.
\end{equation}
It is important to keep in mind that the resolution of identity on the
projectile side is achieved by full basis of states in the Hilbert space of
the projectile and not only by those states that can be obtained from valence
states evolved to the projectile rapidity. In this sense the incoming states
are of a very special type, as by definition we only consider evolved valence
states. The intermediate states for example include states which have no
valence gluons at all, but contain only soft gluons. Such states do not appear
as initial states for the processes we consider. To make a distinction between
the incoming states and the basis states that span the full Hilbert space, we
denote the former by capital letters while the latter by script letters. Thus
for example the $S$-matrix element between two evolved states will be denoted
by
\begin{equation}\label{PP'}
\Sigma_{Y-Y_0}^{PP'}\,\equiv\, 
\langle P_v|\Omega^{P\, \dagger}_{Y-Y_0}\,\, 
\hat S\,\, 
\Omega^{P}_{Y-Y_0}|P'_v\rangle
\end{equation}
omitting the rapidity index whenever we feel it should not cause confusion.

A similar representation can be developed for the target side, see Appendix B.
However we will not consider the most general observables on the target side.
Instead we will concentrate on processes which are either completely inclusive
over the target degrees of freedom, or in which the target scatters
elastically.  For the processes inclusive over the target, the observable
${\hat{\cal O}}$ does not depend on the target degrees of freedom.  Thus
\begin{equation}\label{targetinclusive}
\langle \hat{\cal O}\rangle_{\rm TI}\,=\,\langle T_{Y_0}|\left(\,\delta^{Pp'}\,-\,\hat\Sigma_{Y-Y_0}^{\dagger\, Pp'}\right)
\,\,\,{\cal O}^{p'p''}
\,\left(\delta^{Pp''}\,-\,\hat\Sigma_{Y-Y_0}^{p''P}\right)\,|T_{Y_0}\rangle\,.
\end{equation}
The external average over target evolved to rapidity $Y_0$ can be done using 
the target weight functional $W^T_{Y_0}$ evolved from zero rapidity with the Hamiltonian $H^{JIMWLK}$.
\beq\label{mainP}
\langle {\cal O}\rangle_{\rm TI}\,=\,\int DS\,W^T_{Y_0}[S]\,\,
\sum_{p'p''}\,\left(\delta^{Pp'}\,-\,\Sigma_{Y-Y_0}^{\dagger\, Pp'}[S]\right)\,\,
\left(\delta^{Pp''}\,-\,\Sigma_{Y-Y_0}^{Pp''}[S]\right)\,\, \, {\cal O}^{p'p''}\,.
\eeq

For the target elastic processes only one intermediate state, namely $|T\rangle$ contributes.
In this case we get
\beq\label{targetelastic}
\langle \hat{\cal O}\rangle_{ \rm TE}\,=\,\langle T_{Y_0}|\left(\,\delta^{Pp'}\,-\,\hat\Sigma_{Y-Y_0}^{\dagger\,Pp'}\right)|T_{Y_0}\rangle
\,\,\,{\cal O}^{p'p''}
\,\langle T_{Y_0}|\left(\delta^{Pp''}\,-\,\hat\Sigma_{Y-Y_0}^{p''P}\right)\,|T_{Y_0}\rangle
\eeq
This expression involves two independent target averages and can be written in terms of the
 target weight functional $W$ as
 \beq\label{mainel}
\langle {\cal O}\rangle_{\rm TE}\,=\,\int DS\int D\bar S\,W^T_{Y_0}[S]\,\,W^T_{Y_0}[\bar S]
\sum_{p'p''}\,\left(\delta^{Pp'}\,-\,\Sigma_{Y-Y_0}^{\dagger\, Pp'}[S]\right)\,\,
\left(\delta^{Pp''}\,-\,\Sigma_{Y-Y_0}^{p''P}[\bar S]\right)\,\, \, {\cal O}^{p'p''}\,.
\eeq
In these equations we have written the integration measure as $DS$ rather than $D\alpha$ as in eqs.(\ref{ss},\ref{hee}). This we do purely for notational simplicity. The actual integration measure is indeed $D\alpha$, but since the integrands that we will encounter depend only on $S$ we allow ourselves this notational shortcut. With this note of caution we will use this notation throughout the paper.

Within the above formulation, the problem of high energy evolution of any observable reduces to
identification of corresponding quantum operator $\hat{\cal O}$ and its matrix element $ {\cal O}^{p'p''}$.

In the rest of this section we will discuss several observables related to elastic and diffractive scattering and their evolution with respect to various rapidity variables on which they depend.
To make clear our notations, superscript over any observable refers to the projectile and subscript to the target. Thus for example $N^{D,Y_P}_E$ denotes the cross section for projectile diffraction with the diffractive interval $Y_P$ with target scattered elastically.
The total rapidity of all processes discussed here is denoted by $Y$ and most of the time we will not indicate it explicitly.

We start our discussion with simple observables which depend only on total rapidity.

\subsection{Elastic Scattering}

\begin{figure}[htbp]
\begin{center}
{\epsfig{file=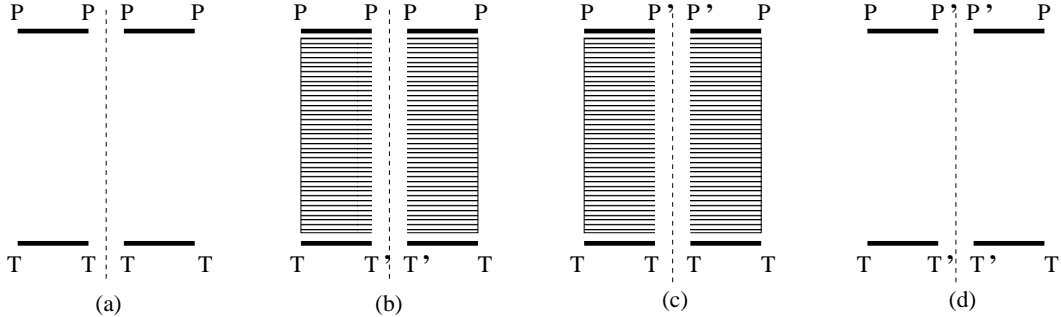,width=140mm}}
\end{center}
\caption{\it Elastic scattering: (a) total elastic: the final states on the cut are the same as the initial ones both on the projectile (P) and target (T) sides; (b)projectile elastic, target inclusive - all final states on the target side $T'$ are summed over; (c) target eslastic, projectile inclusive: all projectile final states $P'$ are summed over; (d)
 double inclusive with maximal
gap. Horizontal lines denote final state gluons.}
\label{fig1}
\end{figure}

\subsubsection{Total elastic scattering}

Consider a totally elastic process where both the projectile and the target scatter elastically Fig. \ref{fig1},a.
For the target averaging we use eq.(\ref{mainel}). On the projectile side the observable $\hat{\cal O}$ is

\beq\label{op}
\hat{\cal O}\,=\,|P_v\rangle\,\langle P_v|\,\,
\eeq
The total elastic cross section reads:
\beq\label{el}
N_{E}^E\,=\,
\left\vert\int DS\,W^T_{Y_0}[S]\,\,\left(1\,-\,\Sigma^{PP}_{Y-Y_0}[S]\right)\right\vert^2\,\,=\,\, T_Y^2\,.
\eeq
This expression obviously does not depend on the rapidity $Y_0$ which separates the projectile and target Hilbert spaces.
The evolution of the amplitude $T_Y=1-{\cal S}_Y$ is given by the JIMWLK equation eq.(\ref{hee}).

\subsubsection{Projectile elastic scattering inclusive over the target}
This observable (Fig. \ref{fig1},b)
is defined as elastic over the valence degrees of freedom of the projectile and inclusive over 
 all the rest of the rapidity interval. Thus we should take the separation rapidity at the total rapidity of the process $Y_0=Y$.
Using eq.(\ref{mainP}) with the operator $\hat{\cal O}$  given by eq.(\ref{op}) we get
\beq\label{el1}
N^E_{I}\,=\,
\int DS\,W^T_{Y}[S]\,\,\left\vert 1\,-\,\Sigma^{PP}_0[S]\right\vert^2\,.
\eeq
The evolution of this observable is given by the JIMWLK evolution of $W^T_Y[S]$.

\subsubsection{Elastic target inclusive over the projectile}

Target valence degrees of freedom scatter elastically while all states on the projectile side over the full rapidity interval are 
summed over inclusively (Fig. \ref{fig1},c). We use eq.(\ref{mainel}) 
\beq\label{eli}
N_{E}^I\,=\,
\int DS\,D\bar S\,W^T_0[S]\,W^T_0[\bar S]\,\,
\left(1\,-\,\Sigma^{PP}_{Y}[S]\,-\,\Sigma^{PP}_{Y}[\bar S^\dagger]\,+\,\Sigma^{PP}_{Y}[\bar S^\dagger\, S]\right)\,
\eeq
We remind the reader that we are interested in the situation where the projectile is dilute and the target is dense all the way through the evolution. This allows us to choose $Y_0=0$ in eq.(\ref{eli}).
The evolution of each $\Sigma$ term in eq.(\ref{eli}) is given by the JIMWLK Hamiltonian which depends on the argument of $\Sigma$. The term of most interest is the last term involving $\Sigma^{PP}_{Y}[\bar S^\dagger\, S]$. 
Using the property of the JIMWLK Hamiltonian eq.(\ref{property}) we can be written 
\beq\label{fel1}
{d\over dY}\,\Sigma^{PP}_{Y}[\bar S^\dagger\, S]\,=\,-H_3[S;\bar S]\,\,\Sigma^{PP}_{Y}[\bar S^\dagger\, S]
\eeq
We have introduced the operator \cite{HWS}
\beq\label{triple}
H_3[S;\bar S]\,=\,\int_z\left[Q^a_i(z,[S])\,+\,Q^a_i(z,[\bar S])\right]^2 \,.
\eeq
The Hamiltonian in this equation can then be interpreted as acting to the left on the target weight functionals. 
Thus the amplitude $N_{E}^I$ can be represented  in the following form
\beq\label{fel2}
N_{E}^I\,=\int DS\,D\bar S\,Z_Y(S,\bar S)\,\,
\left(1\,-\,\Sigma^{PP}_{0}[S]\,-\,\Sigma^{PP}_{0}[\bar S^\dagger]\,+\,\Sigma^{PP}_{0}[\bar S^\dagger\, S]\right)\,
\eeq
with the functional $Z$ satisfying the functional equation
\beq
{d\over dY}Z_Y[S,\bar S]\,=\,-\,H_3[S;\bar S]\,Z_Y[S,\bar S]\,.
\eeq
The initial condition for this evolution is
\beq
Z_{Y=0}[S,\bar S]\,=\,W_0^T[S]\,W_0^T[\bar S]\,.
\eeq

\subsection{Diffraction}
An interesting set of observables is that of diffractive observables with
rapidity gap.  We will be interested in identifying the evolution governing
the diffractive observables both with respect to the total rapidity of the
process $Y$ as well as with respect to the rapidity gap(s).

\begin{figure}[htbp]
\centerline{\epsfig{file=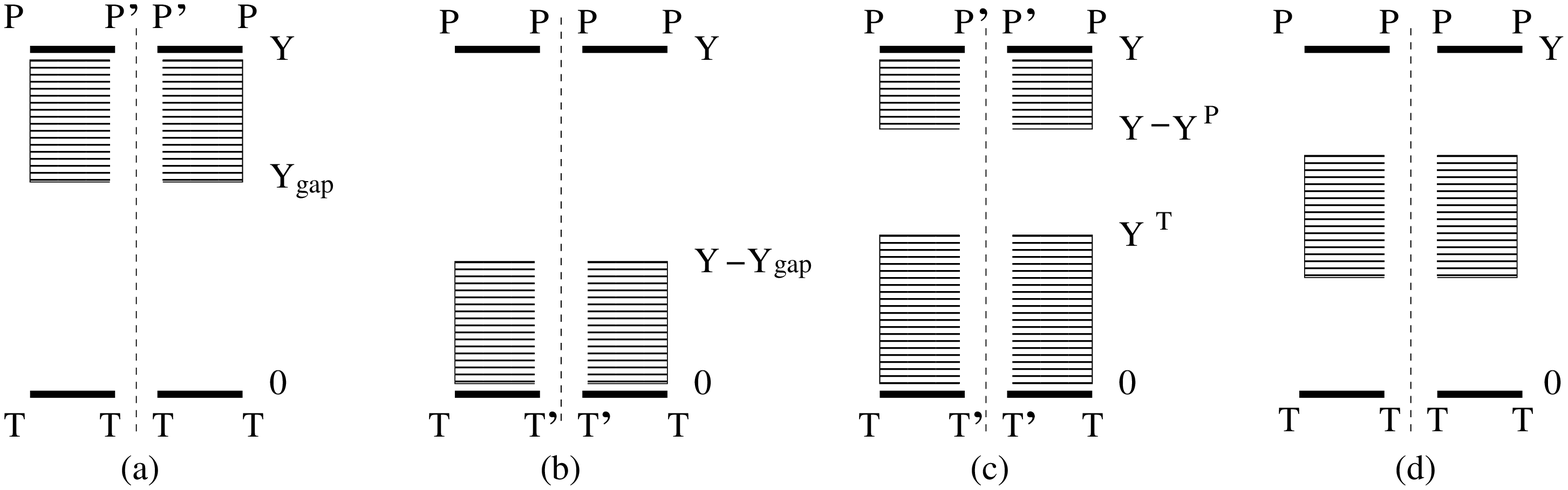,width=120mm}}
\caption{\it Diffraction: (a) Projectile diffraction; (b) Target diffraction; (c) Double diffraction; (d) Central diffraction. Notations are the same as in Fig.1.}
\label{fig2}
\end{figure}

\begin{figure}[htbp]
\centerline{\epsfig{file=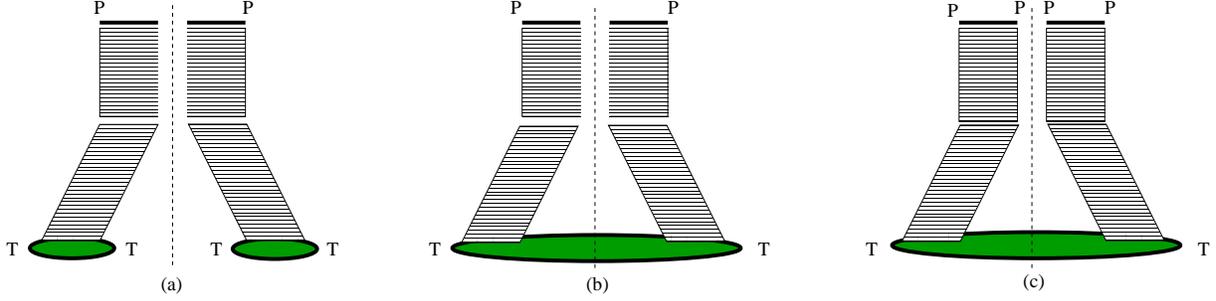,width=160mm}}
\caption{\it Fan diagrams for diffraction: (a) Projectile diffraction with target scattered elastically; 
(b) Projectile diffraction with  target diffracting in a small rapidity interval; (c) Projectile scatters elastically. Notations are the same as in Fig.1.}
\label{fig3}
\end{figure}

\subsubsection{Double inclusive with maximal rapidity gap}
The simplest observable of this type is the cross section 
inclusive  over the valence degrees of freedom of both the target and the projectile 
and with rapidity gap covering the whole rapidity interval $Y$ (Fig. \ref{fig1},d).
We put the separation rapidity $Y_0$ close to the target, $Y_0=0$.
From the target point of view this is an observable of the type eq.(\ref{mainP}). From the point of view of the projectile we have to understand which intermediate states are allowed to contribute in the sum over $p'$. Since we are requiring that no soft gluons be found in the final state, clearly the only states that can contribute are the states of the type $P$, that is projectile states which can be obtained from a valence state by boost to rapidity $Y$. Moreover all such states should be summed over with equal weights. Thus this observable is given by
\beq\label{DI}
N^{D,0}_{D,0}\,=\,\int DS\,W^T_{0}[S]\,\,
\sum_{P'}\,\left(\delta^{PP'}\,-\,\Sigma_{Y}^{\dagger\, PP'}[S]\right)\,\,
\left(\delta^{PP'}\,-\,\Sigma_{Y}^{P'P}[S]\right)\,.
\eeq
The evolution is simply given by the evolution of each factor $\Sigma[S]$ according to the JIMWLK equation. This again can be recast in the form in which the gap as well as the evolution is attributed to the target degrees of freedom. To do this we rewrite the evolution for eq.(\ref{DI}) as follows
\begin{eqnarray}\label{DIev}
{d\over dY}N^{D,0}_{D,0}\,&=&-\,\int DS\,D\bar S\,W^T_{0}[S]\,\,\delta(S-\bar S]\,\times\\
&\times&H_2[S;\bar S]
\sum_{P'}\,\left(\delta^{PP'}\,-\,\Sigma_{Y}^{\dagger\, PP'}[\bar S]\right)\,\,
\left(\delta^{PP'}\,-\,\Sigma_{Y}^{P'P}[S]\right)\,.\nonumber
\end{eqnarray}
with
\beq
H_2[S;\bar S]=H^{JIMWLK}[S]+H^{JIMWLK}[\bar S]
\eeq
Now understanding the action of the evolution operators to the left rather than to the right we can write the same observable as
\beq\label{DI1}
N^{D,0}_{D,0}\,=\,
\int DS\,D\bar S\,\left(1\,-\,\Sigma^{PP}_0[S]\,-\,\Sigma^{PP}_0[\bar S^\dagger]
\,+\, \Sigma^{PP}_0[\bar S^\dagger\, S]\right)\,\,Z_Y[S,\,\bar S]
\eeq
with $Z_Y[S,\,\bar S]$
satisfying the evolution equation
\beq\label{DI11}
{d\over dY}Z[S,\bar S]\,=\,-\,\left(H^{JIMWLK}[S]\,+\,H^{JIMWLK}[\bar S]\right)\,\,Z[S,\bar S]
\eeq
with the initial condition
\beq Z_{Y=0}\,=\,W^T_0[S]\,\delta[S-\bar S]\,.
\eeq
To write eq.(\ref{DI1}) we have used the fact that when the projectile wave function is un-evolved, the states $P'$ constitute complete basis in the projectile Hilbert space, and that the projectile averaged $S$- matrix $\Sigma$ considered as a matrix on the projectile Hilbert space has the property
\beq\label{unit}
\Sigma^{\dagger PP'}_0[S]\,\Sigma^{P'P}_0[\bar S]\,=\,\Sigma^{PP}_0[S^\dagger\bar S]\,.
\eeq

We now consider processes where the projectile diffracts into a rapidity interval $Y_P$. This interval is not necessarily small, so this type of observable can be evolved independently over the total rapidity $Y$ and the width of the diffractive interval $Y_P$. The target can either scatter elastically or can in principle also diffract. We start with the process where the scattering on the target side is elastic, the process considered in \cite{Kovchegov:1999ji}.

\subsubsection{Projectile diffraction with target scattered elastically}
Consider the projectile diffraction with rapidity gap imposed on the target side where the target undergoes elastic scattering
(Figs. \ref{fig2},a and \ref{fig3},a). 
The diffractive interval on the projectile side is $Y_P$ and the rapidity gap $Y_{gap}=Y-Y_P$.
We choose the rapidity separation $Y_0$ to be at the target rapidity $Y_0=0$. We show in Appendix B that the result does not depend on the 
position of the separation rapidity, as it should.
Clearly we should use eq.(\ref{mainel}) on the target side. On the projectile side we have to sum over all intermediate states which can be obtained by the evolution over the gap. Thus the intermediate states $p'$ in eq.(\ref{mainel}) should have the form
\beq\label{ppy}
|p\rangle\,=\,\Omega_{Y_{gap}}\,|p_{Y_P}\rangle
\eeq
where  the states $|p_{Y_P}\rangle$ form a complete basis inside the rapidity interval $Y_P$ and not just on the valence part of the projectile Hilbert space\

In the framework of eq. (\ref{rdo0}) this corresponds to choosing the operator
${\cal O}=\Sigma_{p_{Y_P}}|X\rangle\langle X|$ with $|X\rangle=\Omega^\dagger_{Y_P}\vert p_{Y_P}\rangle$. Note that the states $|X\rangle$ defined through this relation are not necessarily valence states, as remarked in the discussion in Section III.A. In fact, since $\vert p_{Y_P}\rangle$ span the full Hilbert space on the rapidity interval $Y_P$, so do the states $\vert X\rangle$, since $\Omega_{Y_P}$ is a unitary operator. Thus $\Sigma_{\{X\}}\vert X\rangle\langle X\vert=\Sigma_{\{p_{Y_P}\}}|p_{Y_P}\rangle\langle p_{Y_P}|=1_{Y_P}$ where $1_{Y_P}$ is the unit operator inside the rapidity interval $Y_P$.

With this definition of the intermediate states the expression for this observable is
 \beq\label{difel}
N^{D,Y_P}_E\,=\,\int DS\,D\bar S\,W^T_{0}[S]\,\,W^T_{0}[\bar S]
\,\left(1\,-\,\Sigma_{Y}^{PP}[S]-\,\Sigma_{Y}^{\dagger\, PP}[\bar S]+\sum_{p}\Sigma_{Y_{gap}}^{\dagger\, Pp}[Y_P;\bar S]
\,\Sigma_{Y_{gap}}^{pP}[Y_P;S]\right)\,.
\eeq
Here 
\beq
\Sigma_{Y_{gap}}^{Pp}[Y_P;S]\,=\,\langle P_v\vert \Omega^\dagger_{Y_p}\,\Omega^\dagger_{Y_{gap}}\,\hat S\,\Omega_{Y_{gap}}\vert p_{Y_P}\rangle\,.
\eeq
To bring this expression to a simpler form we note that the evolution of $\Sigma_{Y_{gap}}^{Pp}[Y_P;S]$ with respect to $Y_{gap}$ is given by $H^{JIMWLK}$. Thus we can again integrate the evolution by parts in eq.(\ref{difel}) and express 
the integrand in terms of  $\Sigma_{Y_{gap}=0}^{Pp}[Y_P;S]$. However at $Y_{gap}=0$, the states $p$ form a complete basis in the projectile Hilbert space and we can use again the property eq.(\ref{unit}). 
As a result we get the target weight functionals evolved through the gap:
\beq\label{PD1}
N^{D,Y_P}_E\,=\,
\int DS\,D\bar S\,\left(1\,-\,\Sigma^{PP}_{Y_P}[S]\,-\,\Sigma^{PP}_{Y_P}[\bar S^\dagger]
\,+\, \Sigma^{PP}_{Y_P}[\bar S^\dagger\, S]\right)\,\,W^T_{Y_{gap}}[S]\,W^T_{Y_{gap}}[\bar S]\,.
\eeq
This again can be written in terms of the target "weight function" which depends on two unitary matrices and two rapidities
\beq\label{PD2}
N^{D,Y_P}_E\,=\,
\int DS\,D\bar S\,\left(1\,-\,\Sigma^{PP}_{0}[S]\,-\,\Sigma^{PP}_{0}[\bar S^\dagger]
\,+\, \Sigma^{PP}_{0}[S\,\bar S^\dagger]\right)\,\,Z_{Y_P,Y_{gap}}[S,\bar S]\,.
\eeq
The evolution of Z with respect to the width of the diffractive interval as
\beq\label{H3}
{\partial\over \partial Y_P}Z_{Y_P,Y_{gap}}[S,\bar S]\,=\,-\,H_3[S;\bar S]\,\,Z_{Y_P,Y_{gap}}[S,\bar S]
\eeq
with $H_3[S,\bar S]$ defined in eq.(\ref{triple}).
The evolution with respect to the width of the gap is in general more complicated, however for $Y_P=0$ it reduces to
\beq\label{equ}
{\partial\over \partial Y_{gap}}Z_{Y_P=0,Y_{gap}}[S,\bar S]\,=\,-\,
H_2[S;\bar S]\,\,Z_{Y_P=0,Y_{gap}}[S,\bar S]
\eeq
Thus to find $Z_{Y_P,Y_{gap}}$ one has first to evolve $Z$ from $Y_{gap}=0$ to $Y_{gap}$ with eq.(\ref{equ}) starting with the initial condition
\beq\label{H31}
Z_{Y_P=0;Y_{gap}=0}[S,\bar S]\,=\,W^T_0[S]\,\,W^T_0[\bar S]
\eeq
and subsequently evolve the solution with respect to $Y_P$ via eq.(\ref{H3}).
 
This reproduces the result previously derived in Ref. \cite{HWS}.

\subsubsection{Projectile diffraction  with the target diffracting in a small rapidity interval.}
Another interesting diffractive observable is the cross section summed inclusively over the valence target excitations
(Fig. \ref{fig3},b). 
Like before we fix the diffractive rapidity interval on the projectile side and the rapidity gap on the target side, but
 sum inclusively over the possible target valence states. In view of our discussion in the previous subsection the result is easy to write down
 \beq\label{difin}
N^{D,Y_P}_{ D,0}\,=\,\int DS\,W^T_{0}[S]\,\,\left(1\,-\,\Sigma_{Y}^{PP}[S]-\,\Sigma_{Y}^{\dagger\,PP}[S]+\sum_{p}\Sigma_{Y_{gap}}^{\dagger\, Pp}[Y_P;S]\Sigma_{Y_{gap}}^{pP}[Y_P;S]\right)\,\,
\eeq
This can again be rewritten in the form similar to eq.(\ref{PD2})
\beq\label{PD3}
N^{D,Y_P}_{ D,0}\,=\,
\int DS\,D\bar S\,\left(1\,-\,\Sigma^{PP}_{0}[S]\,-\,\Sigma^{PP}_{0}[\bar S^\dagger]
\,+\, \Sigma^{PP}_{0}[\bar S^\dagger\, S]\right)\,\,Z_{Y_P,Y_{gap}}[S,\bar S]\,.
\eeq
The evolution of the functional $Z$  with respect to the width of the diffractive interval is as before
\beq\label{equ3}
{\partial\over \partial Y_P}\,Z_{Y_P,Y_{gap}}[S,\bar S]\,=\,-\,H_3[S;\bar S]\,\,Z_{Y_P,Y_{gap}}[S,\bar S]\,.
\eeq
and its evolution at vanishing $Y_P$ with respect to the rapidity gap is
\beq\label{equ1}
{\partial\over \partial Y_{gap}}Z_{Y_P=0,Y_{gap}}[S,\bar S]\,=\,-\,H_2[S;\bar S]\,\,Z_{Y_P=0,Y_{gap}}[S,\bar S]\,.
\eeq
The only difference relative to the observable in the previous subsection is in the initial conditions. 
One has to solve eq.(\ref{equ1}) with the initial condition
\beq
Z_{Y_P=0;Y_{gap}=0}[S,\bar S]\,=\,W^T_0[S]\,\delta[S-\bar S]
\eeq
and feed the solution as the initial condition into eq.(\ref{equ3}).

\subsubsection{Double diffraction and more.}
In the previous subsection we have considered an observable which is summed over the final states of the target in a small (valence) rapidity interval. The restriction on the size of the diffractive interval on the target side $Y_T$ can be straightforwardly lifted, and we can consider double diffraction with three independent rapidity intervals $Y_P$, $Y_T$ and $Y_{gap}$ denoted by.
Note that in the framework of the Pomeron fan diagrams  (Fig. \ref{fig3}), this observable does not depend on $Y_T$ and $Y_{gap}$ separately but rather only on their sum $Y_T+Y_{gap}$. which is the distance between the valence target and the diffractive remnants of the projectile. This is so 
since the Pomerons can not be cut horizontally without loosing powers of energy. Thus in the absence of Pomeron loops, once a gap is required, no Pomeron can be cut below the gap and no particles are produced. 
We will show in the next section, that this property holds in the dipole model approximation to JIMWLK evolution. However in the full JIMWLK framework this is not the case and double diffraction is indeed possible.

We start with defining the cross section where the projectile states are summed over inclusively over the valence rapidity, target scatters diffractively in the rapidity interval $Y_T$ and require the rapidity gap $Y_{gap}$.
By the same reasoning as in the previous section we have
\beq\label{diftar}
N^{D,0}_{ D,Y_T}\,=\,\int DS\,W^T_{Y_T}[S]\,\,\left(1\,-\,\Sigma_{Y_{gap}}^{PP}[S]-\,\Sigma_{Y_{gap}}^{\dagger PP}[S]\,+\,
\sum_{P'}\Sigma_{Y_{gap}}^{\dagger\, PP'}[S]\,\Sigma_{Y_{gap}}^{P'P}[S]
\right)\,.
\eeq
Note that the sum over $P'$ in the last term does not give unity, since $P'$ do not constitute complete basis of states in the rapidity interval $Y_{gap}$.

Rewriting this in a form similar to the previous subsection we obtain
\beq
N^{D,0}_{ D,Y_T}\,=\,
\int DS\,D\bar S\,\left(1\,-\,\Sigma^{PP}_{0}[S]\,-\,\Sigma^{PP}_{0}[\bar S^\dagger]
\,+\, \Sigma^{PP}_{0}[\bar S^\dagger\, S]\right)\,\,Z_{Y_{gap},Y_T}[S,\bar S]\,.
\eeq
The evolution of the functional $Z$  with respect to the rapidity gap is
given by
\beq\label{equ4}
{\partial\over \partial Y_{gap}}\,Z_{Y_{gap},Y_T}[S,\bar S]\,=\,-\,H_2[S;\bar S]\,\,Z_{Y_{gap},Y_T}[S,\bar S]\,.
\eeq
and its evolution with respect to the diffractive interval $Y_T$ at vanishing $Y_{gap}$ is
\beq\label{equ5}
{\partial\over \partial Y_T}Z_{Y_{gap}=0,Y_T}[S,\bar S]\,=\,-\,H_3[S;\bar S]\,\,Z_{Y_{gap}=0,Y_T}[S,\bar S]\,.
\eeq
To calculate $Z_{Y_{gap},Y_T}$ one has to solve eq.(\ref{equ5}) with the initial condition\footnote{Note that the evolution eq.(\ref{equ5}) of the initial condition eq.(\ref{init1}) is equivalent to the evolution of $W^T[S]$ with the standard JIMWLK Hamiltonian. This follows directly from eq. (\ref{property}).}
\beq\label{init1}
Z_{Y_{gap}=0,Y_T=0}[S,\bar S]\,=\,W^T_0[S]\,\delta[S-\bar S]
\eeq
and then evolve $Z$ with respect to $Y_{gap}$ according to eq.(\ref{equ4}).
The solution can be formally written as 
\beq
Z_{Y_{gap},Y_T}[S,\bar S]\,=\,e^{\,-\,H_2\,Y_{gap}}\,e^{\,-\,H_3\,Y_T}\ \ Z_{Y_{gap}=0,Y_T=0}[S;\bar S]
\eeq

We can now generalize the previous consideration to an observable with arbitrary number of gaps and diffractive intervals. It is clear from the discussion so far that the evolution across any gap is given by the Hamiltonian $H_2$ while the evolution across a diffractive interval is given by $H_3$. Consider the process with $n$ rapidity gaps $Y_1,\ ...\ , \ Y_n$ separated by diffractive intervals $y_0,\ ... \ , y_n$ where $y_0$ is the diffractive interval of the target and $y_n$ that of the projectile. Our previous discussion allows us to write down this observable in the following form:
\beq
N^{D}_{D}(Y_1,...,Y_n;y_0,...,y_n)=\int DS\,D\bar S\,\left(1\,-\,\Sigma^{PP}_{0}[S]\,-\,\Sigma^{PP}_{0}[\bar S^\dagger]
\,+\, \Sigma^{PP}_{0}[\bar S^\dagger\, S]\right)\,\,Z_{\{Y_i\},\{y_j\}}[S,\bar S]\,.
\eeq
where
\beq
Z_{\{Y_i\},\{y_j\}}[S,\bar S]\,=\,e^{\,-\,H_3\,y_n}\ e^{\,-H_2\,Y_n}\,...\,e^{\,-\,H_3\,y_1}\ e^{\,-\,H_2\,Y_1}
\ e^{\,-\,H_3\,y_0\,}\ \ Z_{0}[S;\bar S]
\eeq
with
\beq
Z_{0}[S;\bar S]\,=\,W^T_0[S]\,\delta[S-\bar S]
\eeq

\subsubsection{Connection with the approach of \cite{HWS}.}
We briefly discuss the relation of our approach to that of \cite{HWS}. 
As we have noted above, our results coincide with those of \cite{HWS} where they
overlap.  To calculate the diffractive cross section with the diffractive interval $Y_P$ on the projectile side, the authors of \cite{HWS} introduce the gluon cloud operator in
the form
\begin{equation}
U= {\sf U}_{\text{i}} [\Xi,\xi]{\sf U}_{\text{f}} [\Xi,\xi]
\end{equation} with
\begin{equation}
 {\sf U}_{\text{i}}[\Xi,\xi]
=  
 {{ \sf P}}_{y_2} \exp\Big[ i 
 \int  dy_1dy_2\, \theta(y_1-y_2) b^{ia}_{z}(J_L[S_{y_1}])( 
 S^{ab}_{y_2, z}\Xi^{b,i}_{y_2, z}  
 +
 S^{ab}_{y_2, \bm{z}}  \xi^{b,i}_{y_2, \bm{z}}
 ) 
 \Big]
 \end{equation}
\begin{equation}
 {\sf U}_{\text{f}}[\Xi,\xi]
=  {{ \sf P}}_{y_2} \exp\Big[ i 
 \int  dy_1dy_2\, \theta(y_1-y_2) b^{ia}_{z}(J_L[1])( 
 \Xi^{a,i}_{y_2, \bm{z}}  
 +
  \xi^{a,i}_{y_2, \bm{z}}
 ) 
 \Big]
\ .
\end{equation}
Here $\xi$ and $\Xi$ are both random noise variables. The correlator of $\xi$ is identical with the the vacuum average of free gluon creation and annihilation operators:
$$
\langle \xi^{a, i}_{y_1 \bm{u}}  \xi^{b, j}_{y_2
  \bm{v}}\rangle  =\langle 0|k^+ a^{a, i}(k^+,u)  a^{\dagger b, j}(p^+,v)|0\rangle=
\delta^{ab}\delta^{ij} \delta^{(2)}_{\bm{u},\bm{v}}\delta_{y_1,y_2}
$$
The correlators of $\Xi$ are identical, but only defined in the rapidity interval $Y_P$ close to the projectile.

The scattering amplitude with arbitrary number of gluons in the final state is calculated by averaging $U$ over the random variable $\xi$. The diffractive cross section is then obtained multiplying the amplitude at a given $S$ by the complex conjugate amplitude at $\bar S$ and averaging over the random variable $\Xi$, always contracting one $\Xi$ in the amplitude with one $\Xi$ in the conjugate amplitude. This expression is then averaged over $S$ and $\bar S$ with the same weight functional.

The procedure outlined above parallels exactly our approach described in this section. The equivalent of the cloud operator $U$ in \cite{HWS} is the operator $\Omega_Y^{P\dagger}\hat S\Omega^P_Y$ in eq.(\ref{rdo0}) with the operator $\Omega$ defined in eq.(\ref{co1}). The random noise variables $\xi$ and $\Xi$ emulate the procedure of taking averages of the soft gluon creation and annihilation operators in the valence projectile state. The operator ${\sf U}_{\text{i}}[\Xi,\xi]$ corresponds to our operator $\hat S\Omega^P_Y$, where all the valence charge densities and the soft gluon operators are rotated by the matrix $S$, while the operator 
${\sf U}_{\text{f}}[\Xi,\xi]$ corresponds to $\Omega_Y^{P\dagger}$. As discussed in the previous section the operator $\Omega^P_Y$ is responsible for the emission of soft gluons in the incoming wave function. These gluons scatter eikonally while propagating through the target. This is precisely the role of ${\sf U}_{\text{i}}[\Xi,\xi]$ in the formalism of \cite{HWS}.  On the other hand the rightmost factor $\Omega_Y^{P\dagger}$ in eq.(\ref{rdo0}) is responsible for the final state emissions, which in \cite{HWS} is achieved through the introduction of ${\sf U}_{\text{f}}[\Xi,\xi]$.
The random noise variables $\xi$ reproduce the contributions of the averages of the type $a a^\dagger$, when both the creation and the annihilation operators arise from the expansion of  $\Omega_P$ or $\Omega_P^{\dagger}$ in the same amplitude (to the right of the operator ${\cal O}$ in eq.(\ref{rdo0})) or complex conjugate amplitude (to the left of the operator ${\cal O}$ in eq.(\ref{rdo0})) . The variable $\Xi$ reprodices the contraction whereby $a$ from the conjugate amplitude is contracted with $a^\dagger$ from the amplitude. These contractions are affected by the presence of the operator ${\cal O}$, and for the diffractive cross section exclude the contributions of the gluon operators in the gap.

\subsection{Diffractive observables in the dipole model limit}
In the previous section we have discussed variety of diffractive observables in the JIMWLK/KLWMIJ approach. The evolution of these observables with respect to different rapidities in the process can be expressed in terms of a functional of two variables $Z[S,\bar S]$. The evolution of this functional across rapidity gap is given by the sum of two independent JIMWLK Hamiltonians, while the evolution across intervals where no restriction on final states is imposed is with the mixed JIMWLK Hamiltonian eq.(\ref{triple}). The difference between the various observables is essentially in the initial conditions for the evolution. This is a conceptually simple result. However just like for the case of the total cross section, the structure of the functional evolution is complicated, and in fact even more complicated, since the number of degrees of freedom is doubled.

It is thus useful to understand these observables in the dipole model limit which puts the evolution in terms of simple rather than functional differential equations. Some of the observables have been discussed before but we present the discussion here for completeness.
As discussed in Section 2, the evolution of the cross section of a projectile is very simple. In terms of the cross section at initial rapidity $1-\Sigma_0^{PP}[s]$, the cross section at rapidity $Y$ is given by $1-\Sigma_0^{PP}[s_Y]$ with $s_Y$ satisfying the Kovchegov equation eq.(\ref{kovchegov}). For the elastic cross section  eq.(\ref{el}) the dipole limit
is clearly given by
\beq
N^E_E\,=\,\langle (1\,-\,\Sigma^{PP}_0[s_Y])\rangle_T^2\,=\,T_Y^2
\eeq
where again $s_Y$ is the solution of the Kovchegov equation with the same initial condition as for the total cross section.

\subsubsection{Projectile elastic - target inclusive}
Starting with eq.(\ref{el1}), in the dipole limit we obtain
\beq
N^E_I\,=\,\langle (1\,-\,\Sigma^{PP}_0[ s_Y])^2\rangle_T
\eeq
which for a single dipole projectile becomes also $T^2_Y$. 

Thus in the dipole limit  $N^E_E\,=\,N^E_I$.
This is obviously the consequence of the target factorization approximation.
\eq{el1} involves target average of the product of two identical dipoles sitting on top of each other in the transverse plane. As discussed in Section 2 this is the situation in which we expect the assumption  of the independent scattering of the two dipoles to be maximally violated.
For possible phenomenological application it is therefore worthwhile to relax the target mean field approximation by introducing as an independent degree of freedom the correlator of two dipoles ($\langle s(x,y)s(u,v)\rangle_T$). Note that when the two dipoles are identical this is the same as the scattering cross section of a  gluonic dipole. This correlator
probes target field fluctuations as discussed in Refs. \cite{LL1,LL2,JP,kl1}.  
Following  proposal of Ref. \cite{kl1} one can construct a Gaussian distribution for target weight 
functional. The mean value of the Gaussian would determine $\langle s\rangle_T$ while the variance
would be given by $\langle s^2\rangle_T$.
Within this setup we would have 
\beq
N^E_I\,=\langle \,(1\,-\,\Sigma^{PP}_0[ s_Y])^2\rangle_T
\eeq
where the averaging is performed over the initial Gaussian distribution which characterizes the target\cite{nestor}.
The variance can be fixed by the data on the elastic scattering at lower energy.

\subsubsection{Elastic target scattering inclusive over the projectile}
We refer to eq.(\ref{eli}). Since the averaging over $S$ and $\bar S$ factorizes, it is easy to see that in the large $N_C$ limit the "composite" dipole made of $\bar S^\dagger S$ factorizes into the product of  two "elementary" dipoles 
\beq
{1\over N_C}\langle {\rm tr} [\bar S^\dagger(x)S(x)S^\dagger(y)\bar S(y)]\rangle_T\,=\,
\langle {1\over N_C}{\rm tr}[ S(x)S^\dagger(y)]\rangle_T\,\,\langle {1\over N_C}{\rm tr}[\bar S(y)\bar S^\dagger(x)]\rangle_T
\eeq
Since each target weight function evolves according to JIMWLK equation, this means that the "elementary" dipoles evolve according to the Kovchegov equation.
Thus in the dipole model limit we have
\beq
N^I_E\,=\, 2\,T_Y\,-\,1\,+\,\Sigma^{PP}_0[s^2_Y]
\eeq
with $s_Y$ satisfying the Kovchegov equation eq.(\ref{kovchegov}) with the initial condition
\beq s_{Y=0}\,=\,s\,.
\eeq

\subsubsection{Projectile diffraction with elastic target scattering}
This observable in the dipole limit has been discussed by Kovchegov and Levin \cite{Kovchegov:1999ji}. Consider eq.(\ref{PD1}). In the dipole limit the evolution of $\Sigma^{PP}$ with respect to the diffractive interval at fixed $Y_{gap}$ is given by the simple dipole evolution. Moreover at $Y_P=0$ this observable reduces to the one discussed in the previous subsection. Thus to obtain $N^{D,Y_P}$ at total rapidity $Y$ we can start evolution at $Y_P=0$, evolve over $Y_{gap}$ to get the observable in the previous subsection and subsequently evolve it according to the dipole model over $Y_P$. Thus we have
\beq\label{FPD1} 
N^{D,Y_P}_E\,=\,2\,T_Y\,-\,1\,+\,\Sigma^{PP}_0[s^{el}_{Y_P,Y_{gap}}]
\eeq
where $s^{el}_{Y_P,Y_{gap}}$ is obtained by solving the Kovchegov equation with respect to $Y_P$ with the initial condition
\beq 
s^{el}_{Y_P=0,Y_{gap}}\,=\,s_{Y_{gap}}^2\,.
\eeq
The derivation in \cite{Kovchegov:1999ji} is given for the projectile which is a single dipole.
\eq{FPD1} is the generalization for an arbitrary initial projectile wave function $\Sigma^{PP}_0[s]$.
Diffraction via the  Kovchegov-Levin equation was extensively investigated in Ref. \cite{LLdiff}.

\subsubsection{Projectile diffraction inclusive over the target}
This observable is similar to the one discussed in the previous subsection. The relation between the two is the same as between the totally elastic scattering and projectile elastic, target inclusive cross section. Indeed, examining eqs.(\ref{DI1},\ref{DI11}) we see that 
 \beq\label{FPD12} 
N^{D,Y_P}_I\,=\,2\,T_Y\,-\,1\,+\,\langle\Sigma^{PP}[s^{in}_{Y_P,Y_{gap}}]\rangle_T
\eeq
Here 
\beq\label{aver}
s^{in}_{Y_P=0,Y_{gap}}\,=\,s_{Y_{gap}}^2
\eeq
and the subsequent evolution of $s^{in}_{Y_P,Y_{gap}}$ over $Y_P$ is according to the Kovchegov equation. Thus clearly if we assume target mean field approximation for the averaging over the target in eq.(\ref{FPD12}) we return to the Kovchegov-Levin observable of the previous subsection. However the mean field approximation for eq.(\ref{FPD12}) is maximally violated, as we have to average products of at least two dipole operators at the same point.
The proper way of calculating this observable therefore is again to have an ensemble of configurations for the target averaging. The evolution then has to be performed for each element of the ensemble. For each element of the ensemble we will obtain the analog of Kovchegov-Levin $N^{D,Y_P}_E[s]$ which then has to be averaged over the target ensemble. In this sense we have
\beq
N^{D,Y_P}_I=\langle N^{D,Y_P}_E\rangle_T
\eeq
and we expect the target ensemble averaging to give important corrections.

 \subsubsection{Double diffraction.}
In this section we discuss the fate of the double diffractive cross section in the dipole model limit. We start by considering eq.(\ref{diftar}). 
Our first observation is that for any finite gap only color singlet intermediate states contribute in the sum over $P'$ in eq.(\ref{diftar}) if the initial state is a color singlet.
The physical reason for this is very simple. Recall the definition
\beq\label{again}
\Sigma_{Y_{gap}}^{PP'}\,\equiv\, \langle P_v|\Omega^{P\dagger}_{Y_{gap}}\,\, \hat S\,\, \Omega^{P}_{Y_{gap}}|P'_v\rangle\,.
\eeq
Let us suppose that $|P_v\rangle$ is a physical color singlet state localized in the impact parameter plane. As discussed in detail in \cite{froissart} the JIMWLK evolution has the property that the wave function of the state spreads in the impact parameter plane.  However if the state is color singlet this spread is rather mild - the long distance tails that are generated by the evolution decrease as ${1\over x^2}$. Thus the probability density to find partons in such an evolved state decreases towards the periphery as ${1\over x^4}$. Thus after the evolution the state $|P_v\rangle$ is still localized with all the probability concentrated at central impact parameters. On the other hand for a color nonsinglet state 
$|P'_v\rangle$ the situation is radically different. The Coulomb tail generated by the evolution decreases only as ${1\over|x|}$ and the probability density decreases only as ${1\over x^2}$. Thus  after any finite evolution interval all the probability for such a state is concentrated at spatial infinity. It thus follows immediately that an overlap of an evolved singlet state and an evolved colored state vanishes no matter how small the evolution interval is. The presence of the $\hat S$-matrix operator in eq.(\ref{again}) does not affect this conclusion, since the action of $\hat S$ is completely local in the transverse plane. 

It is easy to put this argument into more technical terms. Acting by the JIMWLK Hamiltonian on the $S$-matrix element of eq.(\ref{again}) we find that the infrared divergences do not cancel in the virtual part. This (negative) divergence 
exponentiates for a finite evolution interval and the matrix element vanishes. In Appendix C we present this calculation for a matrix element between a singlet and an octet dipoles in the dipole model approximation.

We conclude that as long as $Y_{gap}\ne 0$, the colored intermediate states have to be omitted from eq.(\ref{diftar}).
\begin{eqnarray}\label{diftar2}
N^{D,0}_{ D,Y_T}\,&=&\,\int DS\,W^T_{Y_T}[S]\,\,\left(1\,-\,\Sigma_{Y_{gap}}^{PP}[S]\,-\,\Sigma_{Y_{gap}}^{\dagger\,PP}[S]
\,+\, \sum_{{\rm singlet}\,  P'}\Sigma_{Y_{gap}}^{\dagger\, PP'}[S]\ \Sigma_{Y_{gap}}^{P'P}[S]\right) \nonumber\\
&+&\sum_{{\rm colored}\, P'}\int DS\,W^T_{Y_T}[S]\ \Sigma_{0}^{\dagger\, PP'}[S]\ \Sigma_0^{P'P}[S]\ \delta(Y_{gap})\,.
\end{eqnarray}
The form eq.(\ref{diftar2}) is suitable for taking the dipole limit since all the elements in it are singlets, and therefore can be taken to depend on the dipole degree of freedom only, $\Sigma^{PP'}[S]=\Sigma^{PP'}[s]$.

Let us consider the evolution of the cross section with respect to the rapidity gap. Since each factor $\Sigma$ evolves with the dipole Hamiltonian eq.(\ref{chidip}), and since the dipole Hamiltonian is {\it first order} in the derivative with respect to $s$, we have
\begin{equation}\label{diftar3}
{\partial\over\partial Y_{gap}}N^{D,0}_{ D,Y_T}\,=-\,\int DS\,W^T_{Y_T}[s]\,\,H^{dipole}[s]\left(1\,-\,\Sigma_{Y_{gap}}^{PP}[s]\,-\,\Sigma_{Y_{gap}}^{\dagger\,PP}[s]
\,+\, \sum_{{\rm singlet}\,  P'}\Sigma_{Y_{gap}}^{\dagger\, PP'}[s]\ \Sigma_{Y_{gap}}^{P'P}[s]\right)
\end{equation}
Note that the dipole Hamiltonian is Hermitian with respect to the proper integration measure $DS$, and therefore we can integrate the Hamiltonian by parts and put it on $W^T$. We thus have
\begin{eqnarray}\label{diftar4}
{\partial\over\partial Y_{gap}}N^{D,0}_{ D,Y_T}\,&=&-\,\int DS\,H^{dipole}[s]\,W^T_{Y_T}[s]\,\left(1\,-\,\Sigma_{Y_{gap}}^{PP}[s]\,-\,\Sigma_{Y_{gap}}^{\dagger\,PP}[s]
\,+\, \sum_{{\rm singlet}\,  P'}\Sigma_{Y_{gap}}^{\dagger\, PP'}[s]\ \Sigma_{Y_{gap}}^{P'P}[s]\right)\nonumber\\ &=&{\partial\over\partial Y_T}N^{D,0}_{ D,Y_T}
\end{eqnarray}
We see therefore that evolving the cross section with respect to the rapidity gap is the same as evolving it with respect to the target diffractive interval. This establishes that the cross section does not depend separately on $Y_{gap}$ and $Y_T$, but only on the sum $Y_{gap}+Y_T$.
 
The width of the projectile diffractive interval is not essential for the argument. Clearly, we can equally well allow the projectile to diffract in any finite rapidity interval $Y_P$. The cross section   
then depends separately on $Y_P$ and $Y_{gap}+Y_T$.
 Thus the double diffraction in the dipole model is equal to the single diffraction.
 It is also clear that imposing further gaps and/or diffractive intervals at intermediate rapidity does not change the result. The diffractive cross section depends only on two rapidity variables: the diffractive interval of the projectile
 and the total distance in rapidity between the diffractive remnants of the projectile and the valence rapidity of the target.
 
 To restate our conclusion, we find that to define diffractive scattering within the dipole approximation it is sufficient to sum over color singlet states in some rapidity interval $Y_P$ on the projectile side and require an arbitrarily small gap below this interval. This automatically ensures that there are no gluon emissions on the target side of the gap. We note that in \cite{urs} the diffractive scattering was defined indeed simply by summing over color singlet intermediate states on the projectile side.

Note however that the argument does not extend beyond the dipole model, or rather beyond the leading $N_c$ approximation. It was crucial for our proof
in eq.(\ref{diftar4}) that we could represent the evolution with respect to the width of the gap as the single Hamiltonian acting on the sum of the projectile averaged $S$ - matrices. This holds in the large $N_c$ limit, as the Hamiltonian is linear in the functional derivative. However the full JIMWLK Hamiltonian is a quadratic functional of the functional derivative with respect to $S$. Thus the evolution of the last term in the first line in eq.(\ref{diftar2}) can not be represented as the JIMWLK Hamiltonian acting on the product $\Sigma^{\dagger PP'}[S]\Sigma^{P'P}[S]$
even if the states $P'$ are color singlets. Consequently the evolution with respect to the gap cannot be traded for evolution with respect to $Y_T$. We conclude that the subleading in $1/N_c$ terms in the JIMWLK Hamiltonian are responsible for the double and multiple diffraction processes discussed in the previous section.

 This concludes our discussion of diffractive processes. We now turn to examples of other types of observables.

\section{Scattering with momentum transfer}

In this section we consider scattering processes with fixed transverse momentum transfer. All of the observables considered here are inclusive with respect to the final states of the target. 

A wave function of a probe with definite transverse momentum ${\bf p}$ can be written as:
\beq
|P_{\bf p}\rangle\,=\,\Psi_{\bf p}({\bf r})\,=\,\int d^2b\,e^{i\,{\bf p\,b}}\,\Psi(x)
\eeq
where ${\bf b}=\sum_i {\bf x}_i/n$ is the impact parameter of the configuration of $n$ gluons at transverse positions ${\bf x}_i$, and
${\bf r}_i$ denote the relative distances between the gluons.  

\subsection{Elastic scattering with momentum transfer}

We take the initial projectile state to have transverse momentum zero
and the out state to be the same state but with transverse momentum $q$. 
The operator observable that is being measured has the following form:
\beq
\hat{\cal O}^{E,q}\,=\,\,{1\over (2\,\pi)^2}\,
|P_{q}\rangle \,\langle P_{q}|\,=\,{1\over (2\,\pi)^2}\,
\int \,d^2b\,d^2\bar b\,\,e^{i\,(q)\,(b\,-\,\bar b)}\,\,|P_b\rangle \,\langle P_{\bar b}|\,.
\eeq
This corresponds to putting the separation scale $Y_0=Y$, that is the momentum transfer is fixed for the valence part of the projectile wave function.
 For the expectation value of the  observable we obtain 
\beq\label{MT}
N^{E,q}\,=\,\,{1\over (2\,\pi)^2}\,\int DS\,W^T_{Y}[S]\,\,\int d^2b\,d^2v\,e^{i\,v\,q}\,\,
\left(1\,-\,\Sigma_0^{\dagger\, PP}[S(x)]\right)\,\,
\left(1\,-\,\Sigma_0^{PP}[S(x\,+\,v)]\right)\,.
\eeq
 The $Y$ evolution of $N^{E,q}$ is given through the evolution of $W^T$ with the Hamiltonian $H^{JIMWLK}$.

In the dipole limit, assuming the target mean field approximation  we have 
\beq
N^{E,q}(Y)\,=\,\,{1\over (2\,\pi)^2}\,N^2(q,Y)
\eeq
with 
\beq
N(q,Y)\,=\,\int d^2 b\,\, (1\,-\,\Sigma^{PP}_0[s_Y,b]) \,\,e^{i\,b\,q}\,.
\eeq
Since the dipoles in the final state are displaced with respect to the dipoles in the initial state, the mean field approximation is not suspect in this case.

\subsection{Total cross section with momentum transfer}
We now sum inclusively over the final states of the projectile with momentum transfer $q$.
\beq\label{MT3}
N^{I,q}\,=\,\,{1\over (2\,\pi)^2}\,\int DS\,W^T_{Y}[S]\,\,\int d^2b\,d^2v\,e^{i\,v\,q}\,\sum_{P'}\,
\left(\delta^{PP'}\,-\,\Sigma^{\dagger\, PP'}_0[S(x)]\right)\,\,
\left(\delta^{PP'}\,-\,\Sigma^{P'P}_0[S(x\,+\,v)]\right)
\eeq
where the sum over $P'$ runs over all final states of the projectile. 
 The $Y$ evolution of $N^{I,q}$ is again given by the evolution of $W^T$ with $H^{JIMWLK}$.

Let us now consider this variable in the large $N_C$ limit. It is easy to see that even in the large $N_C$ limit and even assuming that the incoming state $P$ contains only singlet dipoles, the observable eq.(\ref{MT3}) can not be calculated without introducing quadrupoles.
Consider for example a non-forward scattering of a single quark dipole. There are two intermediate states that contribute to the scattering, the color singlet and the color octet. For the color singlet state as usual we have 
\beq\Sigma^{PP}_0(x,y)\,={1\over N_C}\,tr [S_F(x)\,S_F^\dagger(y)]
\eeq 
For the (normalized) color octet intermediate state $P'$ we have (see Appendix C)
\beq \Sigma^{P8}_0\,= \,\sqrt{2\over N_C}\,tr[S_F(x)\,\tau^a\,S_F^\dagger(y)]\,.
\eeq
Summing over the singlet and octet states with equal weights we find
\beq\label{MT11}
\sigma(r,q;Y)\,=\,\int DS\,W^T_{Y}[S]\,\,\int {d^2b\,d^2v\over (2\,\pi)^2\,N_C}\,e^{i\,v\,q}\,
tr\left[\left(1\,-\,S_F(x)\,S^\dagger_F(y)\right)\,
\left(1\,-\,S_F({x+v})\,S^\dagger_F({y+v})\right)\right]
\eeq
or
\beq
\sigma(r,q;Y)\,=\,2\, T_Y\,\delta^2(q)\,-\,\,{1\over (2\,\pi)^2}\,\int d^2b\,d^2v\,e^{i\,v\,q}\,\,\left(1\,-\,q(x,y,x+v,y+v)\right)
\eeq
with
\beq
q(x,y,x+v,y+v)\,=\,{1\over N_C}\,tr\left[S_F(x)\,S^\dagger_F(y)\,S_F({y+v})\,S_F({x+v})^\dagger\right]\,.
\eeq
The last term involves the quadrupole scattering probability and is not suppressed by powers of ${1\over N_C}$ 
relative to the first term.
In fact it is easy to see that for an arbitrary projectile made of dipoles only we have
\beq\label{MT31}
N^{I,q}\,=\,2\,T_Y\,\delta^2(q)\,-\,\,{1\over (2\,\pi)^2}\,\int DS\,W^T_{Y}[S]\,\,\int d^2b\,d^2v\,e^{i\,v\,q}(1\,-\,\Sigma^{PP}_0[q(x,y,x+v,y+v)])
\eeq
where $T$ is the total cross section.
Thus we see that in the large $N_C$ limit it is not sufficient to specify the average of the dipole amplitude in the target wave function, but one also needs to specify the quadrupole amplitude.
The JIMWLK evolution of $W^T$ in (\ref{MT31}) can be integrated onto $\Sigma [q]$. As discussed in 
Section 2 the evolution of $\Sigma^{PP}$ is then
\beq\label{111}
\Sigma^{PP}_Y[q]\,=\,\Sigma^{PP}_0[q_Y]
\eeq
where $q_Y$ is the solution of the quadrupole evolution equation (\ref{sq}) derived in Appendix A with the initial 
condition $q_{Y=0}=q$.

\subsection{Total momentum transfer within a fixed rapidity interval.}
In defining this observable we require that the total momentum $q$ is transferred inside the rapidity interval $Y_P$ on the projectile side.
The generalization of eq.(\ref{MT3}) is straightforward
\begin{eqnarray}\label{MT15}
N^{I,Y_P,q}\,&=&\,\int DS\,W^T_{Y-Y_P}[S]\,\int {d^2b\,d^2v\over (2\,\pi)^2}
\,e^{i\,v\,q}\,\sum_{p'}\,
\left(\delta^{Pp'}\,-\,\Sigma_{Y_P}^{\dagger\, Pp'}[S(x)]\right)\,\,
\left(\delta^{Pp'}\,-\,\Sigma^{p'P}_{Y_P}[S(x\,+\,v)]\right)\nonumber\\
&=&\,2\,T_Y\,\delta^2(q)\,-\,\,{1\over (2\,\pi)^2}\,\int DS\,W^T_{Y-Y_P}[S]\int d^2b\,d^2v\,e^{i\,v\,q}\,\left(1\,-\,\Sigma_{Y_P}^{PP}[ S^\dagger(x)S(x+v)]\right)
\end{eqnarray}
This variable can be evolved both with respect to the total rapidity $Y$ and the rapidity interval $Y_P$. A convenient representation for this purpose is
\beq
N^{I,Y_P,q}\,=\,2\,T_Y\,\delta^2(q)\,-\,\,{1\over (2\,\pi)^2}\,\int DS\,D\bar S\,\int d^2b\,d^2v\,e^{i\,v\,q}\,Z_{Y,Y_P}[S,\bar S]\,\left(1\,-\,
\Sigma_0^{PP}[ \bar S^\dagger(x)\,S(x)]\right)\,.
\eeq
The weight functional $Z_{Y,Y_P}$ is found by solving
\beq
{\partial\over \partial Y_P}\,Z_{Y,Y_P}[S,\bar S]\,=\,-\,
H_3[S;\bar S]\,\,Z_{Y,Y_P}[S,\bar S]
\eeq
with the partial derivative taken at fixed $Y-Y_P$ and the initial condition for the evolution
\beq
Z_{Y-Y_P,Y_P=0}[S,\bar S]\,=\,W_{Y-Y_P}^T[S]\,\,\delta[\bar S(x)-S(x-v)]
\eeq
with $W_{Y-Y_P}^T[S]$ evolving according to the JIMWLK equation.

\subsection{Diffraction with momentum transfer}
The simplest observable of this type is elastic projectile scattering with total momentum transfer $q$ and rapidity gap $Y_P$. For this observable we have
 \beq\label{MT1}
N^{E,q,Y_{gap}}\,=\,\int DS\,W^T_{Y-Y_{gap}}[S]\,\,\int {d^2b\,d^2v\over (2\,\pi)^2}\,e^{i\,v\,q}\,\,
\left(1\,-\,\Sigma_{Y_{gap}}^{\dagger\, PP}[S(x)]\right)\,\,
\left(1\,-\,\Sigma_{Y_{gap}}^{PP}[S(x\,+\,v)]\right)
\eeq
The evolution of this observable with respect to total rapidity $Y$ at fixed $Y_{gap}$ is still given by the JIMWLK evolution of $W^T$.

Finally we can ask for total momentum transfer in a diffractive process where the projectile diffracts into the rapidity interval $Y_P$. Combining the results of Section 3 with the earlier discussion in this section we can write
\beq\label{difint}
N^{D,q,Y_P}_{ D,0}\,=\,2\,
T_Y\,\delta^2(q)-\,\int DS\,W^T_{0}[S]\,\int {d^2b\,d^2v\over (2\,\pi)^2}\,
e^{i\,v\,q}\,\,\left(1\,-\,\sum_{p}\Sigma_{Y_{gap}}^{\dagger\, Pp}[Y_P;S(x)]\,\,
\Sigma_{Y_{gap}}^{pP}[Y_P;S(x+v)]\right)\,\,
\eeq
and the summation has exactly the same meaning as for the diffractive process discussed in Section 3.3.2 and 3.3.3.
Following the same logic as in Section 3.3.3 we can rewrite it as
\beq\label{PD35}
N^{D,q,Y_P}_{ D,0}\,=\,2\,T_Y\,\delta^2(q)\,-\,{1\over (2\,\pi)^2}\,\int DS\,D\bar S\,Z_{Y_P,Y_{gap}}[S,\bar S]\,\int d^2b\,d^2v\,e^{i\,v\,q}\,\left(1\,-\,\Sigma_0^{PP}[ \bar S^\dagger(x)S(x)]\right)\,.
\eeq
And the evolution of the functional $Z$ as before is
\beq
{\partial\over \partial Y_P}Z_{Y_P,Y_{gap}}[S,\bar S]\,=\,-\,
H_3[S;\bar S]\,\,Z_{Y_P,Y_{gap}}[S,\bar S]
\eeq
and with respect to the rapidity gap as
\beq
{\partial\over \partial Y_{gap}}Z_{Y_P=0,Y_{gap}}[S,\bar S]\,=\,-\,H_2[S;\bar S]\,\,Z_{Y_P=0,Y_{gap}}[S,\bar S]\,.
\eeq
Like in Section 3.3.3 the initial condition for the evolution with respect to $Y_{gap}$ is
\beq
Z_{Y_P=0;Y_{gap}=0}[S,\bar S]\,=\,W^T_0[S]\,\delta[\bar S(x)-S(x-v)]
\eeq
This have to be evolved first with respect to the gap to $Y_{gap}$ and subsequently with respect to the diffractive interval to $Y_P$.

To get the large $N_C$ limit for this observable we have to understand the evolution from the point of view of
 $\Sigma^{PP}$ rather than $Z$. At zero gap we simply have the observable of the previous subsection
and we need to know the form of $\Sigma_{Y_P}^{PP}$.
This is clearly obtained by
\beq\label{222}
\Sigma^{PP}_{Y_P}\,=\,\Sigma^{PP}_0[s^{Dq}_{Y_P}]
\eeq
with $s^{Dq}$ solving the Kovchegov equation with the initial condition
\beq
s^{Dq}_{Y_P=0}(x,y)\,=\,q(x,y,x+v,y+v)\,.
\eeq
Note the difference between \eq{111} and \eq{222}. In \eq{111} the evolution is that of the quadrupole while
in \eq{222} the dipole evolution with the quadrupole entering as initial condition only. 
The subsequent evolution across the gap is with two independent JIMWLK Hamiltonians with respect to $S$ and $\bar S$. We also know that the octet states do not make it across the gap in the large $N_c$ limit. This means that for the purpose of the $Y_{gap}$ evolution we can write  
\beq
q(x,y,x+v,y+v)\,=\,s(x,y)\,s(x+v,y+v)
\eeq
and evolve each $s$ according to the Kovchegov equation. All said and done,
the dipole model observable is obtained as the target average of
\beq
N^{D,q,Y_P}_{ D,0}\,=\,2\,T_Y\,\delta^2(q)\,-\,\,{1\over (2\,\pi)^2}\,\int d^2b\,d^2v\,(1-\Sigma^{PP}_0[s^{Dq}_{Y_P,Y_{gap}}])
\eeq
with $s^{Dq}_{Y_P,Y_{gap}}(x,y)$ evolved with Kovchegov equation with respect to $Y_P$ from the initial condition
$s^{Dq}_{Y_P=0,Y_{gap}}(x,y)=s_{Y_{gap}}(x,y)s_{Y_{gap}}(x+v,y+v)$ with $s_{Y_{gap}}(x,y)$ evolved by the
 Kovchegov equation from the initial condition $s_{Y_{gap}=0}(x,y)=s(x,y)$.

This concludes our discussion of transverse momentum transfer.

\section{Inclusive gluon production}

The last observable we consider is the inclusive gluon production. Within the dipole model this has been discussed in \cite{Kovchegov:2001sc}. In \cite{Baier:2005dv} the inclusive gluon production was calculated without the dipole approximation, but the rapidity evolution although implied was not explicit. Single gluon production was also discussed in Refs. \cite{BGV,Braungluon,Krasnitz,Albacete,Kharzeev,Nikolaev}

We are interested in a differential cross section $d\sigma/dy\,dk^2$
for production of gluon at rapidity $Y_0$ and transverse momentum $k$. 
At this rapidity the observable is given by the expression\cite{Baier:2005dv}: 
\beq\label{Og}
\hat{\cal O}_g\,=\,\frac{d }{d\, y}\,\langle 0_a| \Omega_{y}\,(1\,-\,\hat S)\,
\Omega_{y}^\dagger\,\,\,n(k,y)\,\,\,
\Omega_{y}\,(1\,-\,\hat S^\dagger)\,\Omega_{y}^\dagger|0_a\rangle|_{y=Y_0}
\eeq
\beq
n(k,y)\,=\,\int^{e^y\,\Lambda}_{e^{Y_0}\,\Lambda}\, dk^-\,a^{\dagger\,a}_i(k,k^-)\,a^a_i(k,k^-)
\eeq
The target $W^T$ is evolved with the Hamiltonian $H^{JIMWLK}$ to the rapidity $Y_0$, while the projectile $W^P$ is evolved
with $H^{KLWMIJ}$ to rapidity $Y\,-\,Y_0$
\beq\label{Og0}
{d\sigma\over dY_0\,dk^2}\,=\,\int DS\,W^T_{Y_0}[S]\,\int D\rho_P\, W^P_{Y-Y_0}[\rho_P]\, \hat{\cal O}_g(Y_0)
\eeq

For KLWMIJ/JIMWLK evolution the operator $\Omega_y$ is known explicitly eq.(\ref{co}), and $\hat{\cal O}_g$ can be computed explicitly 
\beq\label{0g11}
\hat{\cal O}_g\,=\,\int \frac{d^2z}{2\,\pi} \,\frac{d^2\bar z}{2\,\pi} \,e^{i\,k(z\,-\,\bar z)}\,\,Q^a_i(z)\,Q^a_i(\bar z)
\eeq

For a complete evolution operator which includes Pomeron loops the observable will depend also on the multigluon production amplitudes $Q_n$ and it is not known at present.

We now present a short derivation of \eq{0g11}.
 From \eq{shift} we have
\beq
a^a_i(k,k^-)\,C_{y}\,=\,C_{y}\,\left(a^a_i(k,k^-)\,-\,{1\over \sqrt{k^-}}\,b^a_i[\rho_P]\right)
\eeq
and
\begin{eqnarray}
\left(a^a_i(k,k^-)\,-\,{1\over \sqrt{k^-}}\,b^a_i[\rho_P]\right)\,
\,(1\,-\,\hat S^\dagger)\,C_{y}^\dagger|0_a\rangle&=&\left(a^a_i(k,k^-)\,-\,{1\over \sqrt{k^-}}\,b^a_i[\rho_P]\right)\,
C_{y}^\dagger|0_a\rangle\,\nonumber \\ -\,
\hat S^\dagger\,\left(S^{ab}\,a^b_i(k,k^-)\,-\,{1\over \sqrt{k^-}}\,b^a_i[S\,\rho_P]\right)\,C_{y}^\dagger|0_a\rangle 
&=&-\,{1\over \sqrt{k^-}}\,\hat S^\dagger\,C_{y}^\dagger\,(S^{ab}\,b^b_i[\rho_P]\,-\,b^a_i[S\,\rho_P])\,|0_a\rangle
\end{eqnarray}
To leading order in the coupling constant the operator $C$ commutes with $b$.
Thus we  obtain 
\beq\label{Og1}
\hat{\cal O}_g\,=\,(S^{ab}\,b^b_i[\rho_P;k]\,-\,b^a_i[S\,\rho_P;k])
\,\,(S^{ac}\,b^c_i[\rho_P;k]\,-\,b^a_i[S\,\rho_P;k])
\eeq
This coincides with \eq{0g11} if equivalently written in coordinate representation
\beq\label{Og12}
\hat{\cal O}_g\,=\,\int_{z,\bar z} \,e^{i\,k(z\,-\,\bar z)}\,\,
(S^{ab}\,b^b_i[\rho_P;z]\,-\,b^a_i[S\,\rho_P;z])\,\,
(S^{ac}\,b^c_{i}[\rho_P;\bar z]\,-\,b^a_{i}[S\,\rho_P;\bar z])
\eeq
If we keep the nonlinear terms in the expression for the "classical field" $b(\rho)$, \eq{Og1} and \eq{Og0} provide a generalization of the results of \cite{Baier:2005dv} to include some non-linear effects in the projectile wavefunction (Pomeron loops). In that case for consistency we have to use the JIMWLK+/KLWMIJ+ evolution rather than JIMWLK/KLWMIJ discussed in the bulk of this paper.

Back in the JIMWLK/KLWMIJ limit the operator $\hat{\cal O}_g$ is:
\beq\label{Og2}
\hat{\cal O}_g\,=\,{\alpha_s\over \pi}\,\int_{z,\bar z} e^{i\,k(z\,-\,\bar z)}\,\int_{x,y} {(z-x)_i\over (z-x)^2}\,
{(\bar z-y)_i\over (\bar z-y)^2}\,\rho^a_P(x)\,\rho^b_P(y)\,\,
\left[S_z\,S^\dagger_{\bar z}\,-\,S_z\,S^\dagger_{y}\,-\,S_x\,S^\dagger_{\bar z}\,+\,\,S_x\,S^\dagger_{y}
\right]^{ab}
\eeq
As has been noted before \cite{Kovchegov:2001sc}, \cite{BGV} the 
operator $\hat{\cal O}_g$ \eq{Og2} respects the projectile-target factorization. For $W^P$ evolved with the KLWMIJ Hamiltonian  the correlator $\langle\rho^a_P(x)\,\rho^b_P(y)\rangle_P$ satisfies the BFKL equation.
Thus for color singlet projectile and target
\begin{eqnarray}\label{Og3}
{d\sigma\over dY_0\,dk^2}&=&{\alpha_s\over \,\pi}\,
\int_{z,\bar z} e^{i\,k(z\,-\,\bar z)}\,\int_{x,y} {(z-x)_i\over (z-x)^2}\,
{(\bar z-y)_i\over (\bar z-y)^2}\nonumber \\
&\times & n^P(x,y;Y-Y_0)\,\,[\langle T_{z,y}\rangle_{Y_0}\,+\,\langle T_{x,\bar z}\rangle_{Y_0}
\,-\,\langle T_{z,\bar z}\rangle_{Y_0}\,-\,\langle T_{x,y}\rangle_{Y_0}]
\end{eqnarray}
with $\langle T _{x,y}\rangle_T= {1\over N_C^2-1}\langle{\rm tr} [S(x)S^\dagger(y)]\rangle_T$ 
standing for a gluonic dipole scattering amplitude and $n$ defined as 
\beq
n^P(x,y;Y-Y_0)\,=\,\int D\rho_P\, W^P_{Y-Y_0}[\rho_P]\,\,\rho^a_P(x)\,\rho^a_P(y)\,.
\eeq
This can be written as
\begin{eqnarray}\label{Og4}
{d\sigma\over dY_0\,dk^2}&=&{\alpha_s\over \,\pi}\,\int_{p,q} {(pq)\over p^2\,q^2}\,
\int_{z,\bar z,x,y} e^{i\,(k+p)z\,+\,(q-k)\bar z\,-\,px\,-\,qy}\,\nonumber \\
&\times & n^P(x,y;Y-Y_0)\,\,[\langle T_{z,y}\rangle_{Y_0}\,+\,\langle T_{x,\bar z}\rangle_{Y_0}
\,-\,\langle T_{z,\bar z}\rangle_{Y_0}\,-\,\langle T_{x,y}\rangle_{Y_0}]
\end{eqnarray}
This expression does not assume the dipole model limit nor is it restricted to an initial state consisting of a single dipole. 
The wave function of the projectile enters only through the initial conditions on the evolution of $n^P$.

In terms of the Fourier transforms
\beq
\bar n_{Y-Y_0}^P(k_1,k_2)\,=\,\int_{x,y} e^{-\,i\,k_1\,x\,-\,i\,k_2\,y}\,n^P(x,y;Y\,-\,Y_0)\,;\,\,\,\, \ \  
\bar T_{Y_0}(k_1,k_2)\,=\,\int_{x,y} e^{-\,i\,k_1\,x\,-\,i\,k_2\,y}\,\langle T_{x,y}\rangle_{Y_0}
\eeq
\eq{Og4} takes the form
\begin{eqnarray}\label{Og41}
{d\sigma\over dY_0\,dk^2}&=&{\alpha_s\over \,\pi}\,\int_{p,q} \,K_{2\rightarrow 2}(k;\,q,\,p)
\,\,
\bar n^P_{Y-Y_0}(p,q)\,\bar T_{Y_0}(-k-p, \,k-q)
\end{eqnarray}
with the  vertex 
\beq
K_{2\rightarrow 2}(k;\,q,\,p)\,=\,\left[ {q_i\over q^2}\,+\,{k_i\over k^2}\right]\,\left[{k_i\over k^2}\,-\, {p_i\over p^2}
\right]
\eeq
If either projectile or target is assumed to have translational invariance in the transverse plane, the constraint $p=-q$ is authomatically imposed.
In this approximation the result reduces to the standard $k_t$ factorized form.

For future applications we note that it is sometimes useful to perform the integral over $\rho^P$ in \eq{Og0}.
This averaging procedure as always turns $W^P[\rho^p]$ into $\Sigma[S]$ and any additional factor of $\rho^P$ present in $\hat{\cal  O}_g$ into right or left SU(N) rotation generators \eq{gen}. 
To this end it is useful to temporarily set the second factor of $S$ in eq.(\ref{Og0}) to $\bar S$. Some algebra then gives
\beq\label{Og0g}
{d\sigma\over dY_0\,dk^2}\,=\,\int_{z,\bar z} \,e^{i\,k(z\,-\,\bar z)}\,\,\int DS\,
D\bar S\,\delta[S-\bar S]\,W^T_{Y_0}[S]\,\,Q^a_i(S,z)\,Q^a_i(\bar S,\bar z) \,\,\Sigma_{Y-Y_0}[\bar S^\dagger S]\,.
\eeq 
This representation is similar to the one used in Sections 3, 4  and 5 for other semi inclusive observables.

\section*{Aknowledgments}
The work of H.W. is supported in part by the U.S. Department of Energy
under Grant No. DE-FG02-05ER41377. The work of A.K. is supported by the Department of Energy
under Grant No. DE-FG02-92ER40716.

\appendix

\section{Appendix: Evolution of the quadrupole operator} \label{sec:A}

In this Appendix we consider an evolution of a quadrupole operator \cite{Mueller+,JMK}
\beq
q_{x,y,u,v}\,=\,\frac{1}{N}\,tr[S_x\,S^\dagger_y\,S_u\,S^\dagger_v]
\eeq
The evolution of $q$ follows from the action of the JIMWLK Hamiltonian on $q$ (second equation of the Balitsky`s
hierarchy  \cite{Balitsky}).
Let us introduce three kernels. The Weiszacker-Williams kernel
\beq
K_{x,y;z}\,=\,{(x\,-\,z)_i\,(y\,-\,z)_i\over (x\,-\,z)^2\,(y\,-\,z)^2}
\eeq
The dipole kernel
\beq
M_{x,y;z}\,=\,K_{x,x,z}\,+\,K_{y,y,z}\,-\,K_{x,y,z}\,-\,K_{y,x,z}\,=\,
{(x\,-\,y)^2\over (x\,-\,z)^2\,(y\,-\,z)^2}
\eeq
The ``$4\rightarrow 4$'' kernel \cite{Mueller+,kl1}
\beq
L_{x,y,u,v;z}\,=\,\left[{(x\,-\,z)_i\over (x\,-\,z)^2}\,-\,{(y\,-\,z)_i\over (y\,-\,z)^2}\right ]\,\,
\left[{(u\,-\,z)_i\over (u\,-\,z)^2}\,-\,{(v\,-\,z)_i\over (v\,-\,z)^2}\right ]
\eeq
To derive the evolution of $q$ we follow the same procedure as for the derivation of Kovchegov equation. Namely we take the evolution equation for four Wilson lines and factor its right hand side using the large $N_C$ factorization. The result is \cite{JMK}
\begin{eqnarray}\label{q}
{d q_{x,y,u,v}\over d Y}&=&\frac{\bar \alpha_s}{2\,\pi}\,\int_z \left\{-\,[M_{x,y;z}\,+\,M_{u,v;z}\,-\,L_{x,u,v,y;z}]
\,\,q_{x,y,u,v}\,-\,
 L_{x,y,u,v;z}\,s_{x,v}\,s_{y,u}\,-\,L_{x,v,u,y;z}\,s_{x,y}\,s_{u,v}\right.\,\nonumber \\
&+&\,L_{x,v,u,v;z}\,q_{x,y,u,z}\,s_{z,v}
\left.+\,L_{x,y,x,v;z}\,q_{z,y,u,v}\,s_{x,z}
\,+\,L_{x,y,u,y;z}\,q_{x,z,u,v}\,s_{z,y}\,+\,L_{u,y,u,v;z}\,q_{x,y,z,v}\,s_{u,z}\right\}
\,\nonumber \\
\end{eqnarray} 
Note that as opposed to the Kovchegov equation, \eq{q} is linear in $q$. It is however coupled to $s$ whose evolution is nonlinear \cite{JMK}.
 In order to 
compute the energy behavior of the quadrupole, one first needs to solve the Kovchegov equation for dipoles, and then
solve the linear equation for the quadrupole coupled to dipoles. Note that due to the inhomogeneous term in \eq{q} even for the initial condition $q=0$ a nonvanishing quadrupole is generated by the evolution.

We now comment on the physical meaning of the quadrupole operator $q$.
Consider a projectile which consists of two dipoles at points $x,y$ and
$u,v$. The normalized wave function of such a projectile (in terms of the
creation and annihilation operators of the "quarks" and "antiquarks") is
\beq\label{sin}
|(xy),(uv)\rangle\,=\,{1\over N_C}\,\sum_i q_i(x)\,\bar q_i(y)\,\sum_j
q_j(u)\,\bar q_j(v)\,|0\rangle\,;
\eeq
where $i,\, j$ are color indices in the fundamental representation.
After propagating through the target fields the state emerges with the
wave function
\beq
|(xy),(uv)\rangle_{out}\,=\,{1\over N_C}\,\sum_i
[S(x)q(x)]_i\,[S^\dagger(y)\bar q(y)]_i\,\sum_j
[S(u)q(u)]_j\,[S^\dagger(v)\bar q(v)]_j\,|0\rangle\,.
\eeq
It is easy to calculate the overlap of this outgoing wave function with
the incoming one.
\beq
\langle(xy),(uv)|(xy),(uv)\rangle_{out}\,=\,s(xy)\,s(uv)\,.
\eeq
One can also calculate the overlap of the outgoing state with the two
dipole state where the quarks exchanged their antiquark partners
\beq
\langle(xv),(uy)|(xy),(uv)\rangle_{out}\,=\,{1\over N_C}\,q(xyuv)\,.
\eeq
Thus the quadrupole operator is the antiquark exchange amplitude.
For the single dipole pair of eq.(\ref{sin}) the quadrupole does not
contribute to the total cross section, or forward scattering amplitude.
However a general color singlet state with two quarks and two antiquarks
is a superposition
of two possible dipole pairs
\beq\label{gen1}
\vert singlet,\, xyuv\rangle=\alpha |(xy),(uv)\rangle+\beta|(xv),(uy)\rangle
 \eeq
 The forward scattering amplitude for such state is
 \beq
 \Sigma=\alpha\alpha^*s(xy)s(uv)+\beta\beta^*s(xv)s(uy)+{1\over
N_C}\left[\alpha^*\beta q(xyuv)+\beta^*\alpha q(xvuy)\right]
 \eeq
 Thus the quadrupole contributes $1/N_C$ correction to the total cross
section.
 In the leading order in $1/N_C$ therefore the quadrupole contribution is
absent. However for a state with $n$ dipoles
 there are $n^2$ dipole pairs which can exchange antiquarks. Thus the
quadrupole contribution can become important (depending on the exact wave
function) already when the number of dipoles $n\sim N_C^{1/2}$. On this
note we also mention that the equation for $q$ contains an inhomogeneous
term. Thus even if $q$ vanishes at initial rapidity it is generated
through the evolution. This in particular means that the dipole model
limit is not recovered from eq.(\ref{quadrupoleev}) by setting $q=0$ but
rather by dropping the $q$ - dependence in the function $\Sigma[s,q]$.

\section{Appendix: Attributing final states to the target Hilbert space} \label{sec:B}

In the body of this paper we have avoided introducing intermediate states in the target Hilbert space. Nevertheless this can be done. In this way we can relate the functional $Z[S,\bar S]$ introduced in Section 3 to the density matrix of the target.

We take the target wave function to be boosted to rapidity $Y$.
To discuss the resolution of identity on the target Hilbert space it is convenient to introduce the basis of eigenstates of the operators $\hat Sy=P\exp\{i\int_{-\infty}^{e^y\Lambda} dx^-T^a\alpha^a(x,x^-)\}$ {\it for all $0\le y\le Y$}:
\beq
\hat S_y\,|\{S_y\}\rangle\,=\,S_y\,|\{S_y\}\rangle\ \ \ \ \ \ \ \ \ \ \ \ \ \ \ \ \ \ \ \ \ 
\hat \Sigma^{PP'}[\hat S]\,|\{S_y\}\rangle\,=\,\Sigma^{PP'}[S_Y]\,|\{S_y\}\rangle
\eeq
Any target state $\langle T_Y|$ which depends on all intermediate rapidities $0\le y\le Y$ can be expended in this basis:
\beq\label{Tbasis}
|T_Y\rangle\,=\,\int DS_y\, \Theta^T_Y[S_y]\,|\{S_y\}\rangle \,; \ \ \   
W^T_Y[S_y]\,\equiv\,\Theta^T_Y[S_y]\,\Theta^{*\,T}_Y[S_y]\,; \ \ \ 
\langle T_Y\,|\,T_Y\rangle\,=\,\int DS_y\,\,W^T_Y[S_y]\,=\,1\,.
\eeq
In this equation the functional integral is over $S_y$ for all $0\le y\le Y$. This is defined simply as the integral for all $x^-$ dependent $\alpha(x^-)$: $ DS_y\equiv \Pi_{x^-}D\alpha^a(x^-)$.

Just like for the projectile off diagonal matrix elements it is easy to see that both $W^T_Y$ as well as the target off-diagonal matrix element 
$ \Theta^T_Y\,\Theta^{*\,T'}_Y$ evolve with the Hamiltonian $H^{JIMWLK}$. This is a consequence of Lorentz invariance
for any element of the $S$-matrix.
Resolving identity on the target side in this basis for an arbitrary observable ${\cal O}$ \eq{rdo} we get:
\begin{eqnarray}\label{main}
\langle {\cal O}\rangle_Y&=&\int DS_y\,D\bar S_{\bar y}
\sum_{p'p''}\left(\delta^{Pp'}\,-\,\Sigma^{\dagger\, Pp'}_{Y-Y_0}[\bar S_y]\right)\,
\left(\delta^{Pp''}\,-\,\Sigma^{p''P}_{Y-Y_0}[ S_{ y}]\right)\,\nonumber \\
&\times& \Theta^T_{Y_0}[S_y]\,\Theta^{*\,T}_{Y_0}[\bar S_{ y}]\, {\cal O}^{p'p''}[S_y,\bar S_{ y}]\,.
\end{eqnarray}

In particular consider an observables which is completely inclusive over the projectile degrees of freedom. This includes projectile diffractive observables when we choose the separation scale $Y_0$ such that it includes the rapidity gap on the target side.
\beq
{\cal O}^{p'p''}[S,\bar S]\,=\,{\cal O}[S,\bar S]\,\,\delta_{p'p''}
\eeq
\eq{main} can be rewritten in the following form
\beq\label{mainT}
\langle {\cal O}\rangle_Y\,=\,\int DS_y\,D\bar S_{ y}\,
\sum_{p'}\left(\delta^{Pp'}\,-\,\Sigma^{\dagger\, Pp'}_{Y-Y_0}[\bar S_y]\right)
\left(\delta^{p'P}\,-\,\Sigma^{p'P}_{Y-Y_0}[ S_{y}]\right)\,
\Theta^T_{Y_0}[S_y]\,\Theta^{*\,T}_{Y_0}[\bar S_{\bar y}]\, {\cal O}[S_y,\bar S_{\bar y}]\,.
\eeq
Taking into account that 
\beq
\sum_{p'}\Sigma^{\dagger\, Pp'}_{Y-Y_0}[\bar S_y] \,\,\Sigma^{p'P}_{Y-Y_0}[ S_{ y}]
\,\,=\,\,\Sigma^{PP}_{Y-Y_0}[\,\bar S^\dagger_{ y}S_y] 
\eeq
we can write
\beq\label{mainT1}
\langle {\cal O}\rangle_Y\,=\,\int DS_y\,D\bar S_{ y}\,\left(1\,-\,\Sigma^{PP}_{Y-Y_0}[S]\,-\,\Sigma^{\dagger\, PP}_{Y-Y_0}[\bar S^\dagger_{y}]
\,+\, \Sigma^{PP}_{Y-Y_0}[\,\bar S^\dagger_{y}S_y]\right)\,\,Z_{Y_0}^T[S_y,\bar S_{ y}]
\eeq
where we have introduced
\beq
Z_{Y_0}^T[S,\bar S]\,\equiv\,\Theta^T_{Y_0}[S]\,\,\Theta^{*\,T}_{Y_0}[\bar S]\,\,\, {\cal O}[S,\bar S]
\eeq
This is in the form \eq{PD2}. Note that the functional $Z$ includes information both on the target and observable. 
For example for elastic target scattering
\beq
{\cal O}[S,\bar S]\,=\,\Theta^{*T}_{Y_0}[S]\,\,\Theta^{T}_{Y_0}[\bar S]
\eeq
and we recover \eq{PD1}.
With respect to $Y_0$ this observable obviously evolves with $H^{JIMWLK}[S]+H^{JIMWLK}[\bar S]$.
On the other hand the evolution of any $Y_0$ independent observable follows
from the known evolution of the projectile: 
\beq
{\partial\over\partial Y}\,Z^T[S,\bar S]\,\,=-\,\,\int_z \,\left(\bar Q_i^a[S]\,\,+\,\,\bar Q_i^a[\bar S]\right)^2\,\,
\,Z^T[S,\bar S]\,.
\eeq
This is precisely the evolution we have found in Section 3. This illustrates that we can put the separation scale $Y_0$ either above or below the rapidity gap with the same results for the evolution, as expected.

\section{Appendix: Infrared divergence of the color octet evolution} \label{sec:C}
In this appendix we derive the evolution equation for the scattering amplitude of a dipole into a color  octet final state.

The normalized singlet and octet states defined via "quark" and "antiquark"
creation operators are 
\beq
|P\rangle\,=\,|singlet\rangle\,=\,{1\over \sqrt N}\,\sum_i q_i(x)\,\bar q_i(y)\,|0\rangle\,; \ \ \ \ \ \ \ \ \ \ \ 
|P'\rangle\,=\,|octet \rangle\,=\,{\sqrt 2}\,\bar q(y)\,\tau^a\, q(x)\,|0\rangle
\eeq
The $S$-matrix element $\Sigma^{PP'}_{xy}$ reads 
\beq
\Sigma^{PP'}_{xy}\,=\,\sqrt{2\over N}\, tr[S^\dagger_F(y)\,\tau^a\,S_F(x)]
\eeq
We act with the JIMWLK Hamiltonian on $\Sigma^{PP'}_{xy}$. Evaluating separately each of three terms in the Hamiltonian we obtain 
\beq
\int_{u,v,z} K_{u,v,z}\,J_R^b(u)\,J_R^b(v)\,\,\Sigma^{PP'}_{xy}\,=\,C_F\,\int_z\,M_{x,y,z}\,\Sigma^{PP'}_{xy}
\eeq
\beq
\int_{u,v,z} K_{u,v,z}\,J_L^b(u)\,J_L^b(v)\,\,\Sigma^{PP'}_{xy}\,=\,C_F\,\int_z [K_{x,x,z}\,+\,K_{y,y,z}]
\,\Sigma^{PP'}_{xy}\,+\, {1\over 2\, N}\,\int_z [K_{x,y,z}\,+\,K_{y,x,z}]
\,\Sigma^{PP'}_{xy}
\eeq
\begin{eqnarray}
2&&\int_{u,v,z} K_{u,v,z}\,J_L^b(u)\,S^{bc}(z)\,J_R^c(v)\,\,\Sigma^{PP'}_{xy}\,=\ \ \ \ \ \ \ \ \ \ \ \ \ \ \nonumber \\
&& \ \ \ \ \ \ \ \ \ \ = N\,\int_z \left\{[K_{x,x,z}\,-\,K_{x,y,z}]
\,\Sigma^{PP'}_{zy}\,s_{x,z}\,+\, [K_{y,y,z}\,-\,K_{y,x,z}]\,\Sigma^{PP'}_{xx}\,s_{z,y}\right\}
\end{eqnarray}
$C_F=(N^2\,-\,1)/2\,N$.
As opposed to the equation for the singlet transition amplitude, there is no cancellation of the  infrared divergencies 
as $z\rightarrow \infty$ between the virtual terms ($J_R\,J_R$ and $J_L\,J_L$) and real term:
\begin{eqnarray}
H^{JIMWLK}\,\Sigma^{PP'}_{xy}&=&-\,\int_z C_F\,[M_{x,y,z}\,+\,K_{x,x,z}\,+\,K_{y,y,z}]
\,\Sigma^{PP'}_{xy}\,+\, {1\over 2\, N}\,\int_z [K_{x,y,z}\,+\,K_{y,x,z}]
\,\Sigma^{PP'}_{xy}\,+\nn \\
&+&N\,\int_z\left\{[K_{x,x,z}\,-\,K_{x,y,z}]
\,\Sigma^{PP'}_{zy}\,s_{x,z}\,+\, [K_{y,y,z}\,-\,K_{y,x,z}]\,\Sigma^{PP'}_{xx}\,s_{z,y}\right\}
\end{eqnarray}
Consequently the total action of the  JIMWLK Hamiltonian on $\Sigma^{PP'}_{xy}$ is divergent and negative as the virtual terms
enter the evolution equation with the negative sign. We thus conclude that $\Sigma^{PP'}_{xy}$ vanishes after evolution
 over arbitrarily small rapidity interval.

\end{document}